\documentclass[useAMS,usenatbib]{mn2e}
\usepackage{pifont}
\usepackage{graphicx}
\usepackage{multirow}
\usepackage{textcomp}
\usepackage{amssymb}
\usepackage{color}
\newcommand{\Msolar}{M$_{\odot}$ }
\newcommand{\Msolarnsp}{M$_{\odot}$}

\newcommand{\Zsolarnsp}{Z$_{\odot}$}
\newcommand{\Simname}{\includegraphics[height=0.3cm]{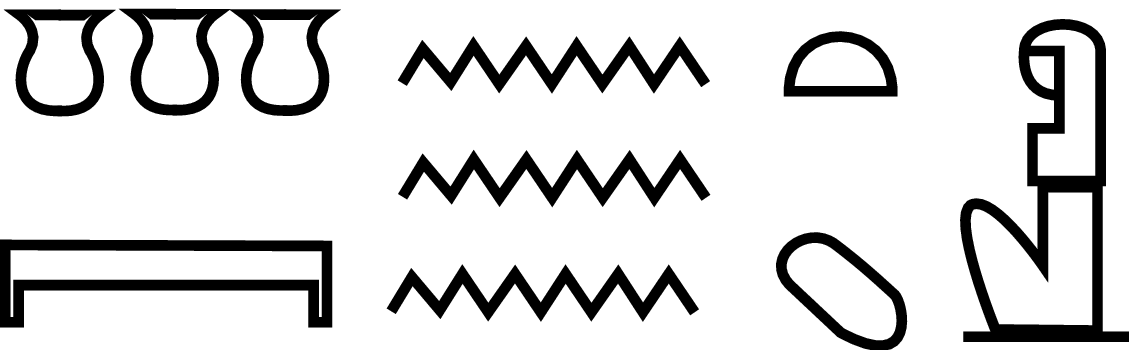} }
\newcommand{\Subvir}{$_{vir}$ }
\newcommand{\Nutref}{\citep{Powell2011,Kimm:2011p2100}}
\newcommand{\AndromedaReferences}{\protect\cite{Collins2011a,Collins:2010p2061,Ferguson:2000p2066,Irwin:2008p2068}; \protect\citet[priv. comm.]{Irwin:2011p2072}; \protect\cite{Kalirai:2010p2065,Letarte:2009p2063,Martin:2009p2070,McConnachie:2008p2069,Morrison:2003p2064,Pustilnik:2008p2067,Richardson:2011p2071} }

\newcommand{\tick}{\ding{51}}

\RequirePackage{ifpdf} 
\begin{document}
\title{Satellite Survival in Highly Resolved Milky Way Class Halos}
\author[S. Geen]
      {Sam Geen$^{1,2}$, Adrianne Slyz$^{1}$, Julien Devriendt$^{1,2}$\\
{$^{1}$University of Oxford, Astrophysics, Keble Road, Oxford OX1 3RH, UK}\\
{$^{2}$CRAL, Universit\'e de Lyon I, CNRS UMR 5574, 
ENS-Lyon,  9 avenue Charles Andr\'e, 69561 Saint-Genis-Laval, France}\\}
\date{\today}
\maketitle

\begin{abstract}
Surprisingly little is known about the origin and evolution of the Milky Way's satellite galaxy companions. UV photoionisation, supernova feedback and interactions with the larger host halo are all thought to play a role in shaping the population of satellites that we observe today, but there is still no consensus as to which of these effects, if any, dominates. In this paper, we revisit the issue by re-simulating a Milky Way class dark matter (DM) halo with unprecedented resolution. Our set of cosmological hydrodynamic Adaptive Mesh Refinement (AMR) simulations, called the Nut suite, allows us to investigate the effect of supernova feedback and UV photoionisation at high redshift with sub-parsec resolution. We subsequently follow the effect of interactions with the Milky Way-like halo using a lower spatial resolution (50pc) version of the simulation down to $z=0$. This latter produces a population of simulated satellites that we compare to the observed satellites of the Milky Way and M31. We find that 
supernova feedback reduces star formation in the least massive satellites but enhances it in the more massive ones. Photoionisation appears to play a very minor role in suppressing star and galaxy formation in all progenitors of satellite halos. By far the largest effect on the satellite population is found to be the mass of the host and whether gas cooling is included in the simulation or not. Indeed, inclusion of gas cooling dramatically reduces the number of satellites captured at high redshift which survive down to $z=0$.
\end{abstract}

\section{Introduction}
\label{introduction}

In the standard cold dark matter paradigm of galaxy formation, galaxies grow inside dark matter halos that merge hierarchically. In other words, smaller halos are captured by larger halos and the galaxies 
they contain become satellite galaxies of the host galaxy until dynamical friction finally forces them to coalesce. Early attempts to reproduce the observed Milky Way satellite population using dark-matter-only simulations overproduced the number of low mass satellites by several orders of magnitude when attributing to each simulated dark matter satellite 
a galaxy using a constant mass-to-light ratio \citep{Moore:1999p565}. Current state-of-the-art dark matter simulations, such as the Aquarius \citep{Springel:2008p979} and the Via Lactea II \citep {Diemand:2008p986} still reach the same conclusion.

Whilst the existing sample of Milky Way satellite galaxies is almost certainly incomplete \citep{Koposov:2008p793,Tollerud:2008p1758} and new satellite galaxies are continually being discovered (e.g. \cite{Belokurov:2010p1746}),  it appears extremely unlikely that new observations will uncover galaxies populating every dark matter substructure predicted to exist around the Milky Way. As a result, various authors have attempted to explain this discrepancy by invoking physical mechanisms that reduce or prevent star formation in the majority of the smaller halos, making these dark matter substructures fainter or completely dark in the process. In particular, photoionisation and feedback from supernovae have been proposed as the most likely mechanisms to prevent gas from condensing to form stars, although other mechanisms, such as cosmic rays, have also been suggested \citep{Wadepuhl2010a}.

The UV ionising background has been argued to be effective at halting or preventing star formation in low-mass halos in studies using analytic arguments, observations, N-body simulations and semi-analytic models \citep{EfstathiouG.1992,Bullock:2000p1416,Benson:2002p1802,Somerville:2002p1760,Kravtsov:2004p1761,Moore:2006p1747,Simon:2007p1453,Strigari:2007p801,Madau:2008p1762,Maccio2010,Munoz2009a,Busha:2010p1756}. Following the pioneering work of \cite{QuinnThomas1996,Gnedin:2000p1832}, hydrodynamic simulations of Milky Way-like galaxies have recently been employed to study the problem \citep{Hoeft:2006p1811,Libeskind:2010p1751,Nickerson:2011p2094,Okamoto:2008p1812,Okamoto:2009p1754,Parry2012,Sawala2012a,Scannapieco2011a,Wadepuhl2010a,Bovill2011}. Many of these authors have found a best fit to the observed luminosity function by adopting instantaneous reionisation at $z \sim 11$, in agreement with the latest WMAP estimate ($z_{reion} = 10.5 \pm 1.2$ \citep{Larson:2010p1748}). However, \cite{Hoeft:2006p1811,
Wadepuhl2010a,
Guo2010a} also find reionisation to have limited effectiveness in completely suppressing star formation in low-mass halos that have already begun forming stars. \cite{Hoeft:2006p1811} determine that there is a characteristic mass of $6.5\times10^9 h^{-1} $\Msolar below which halos become unable to retain baryons down to $z=0$, and hence cannot form stars. \cite{Okamoto:2008p1812} find a similar result, which they translate into a minimum circular velocity v$_{max}$ of 25km/s below which halos are dark. \cite{Okamoto:2009p1754} further refine this finding, stating that  the  cut-off should be lowered to v$_{max}$ = 12km/s at reionisation ($z=9$ in their case). To further obscure the picture, the extent of the epoch of reionisation itself is poorly constrained, with only a lower limit  of $z \gtrsim 6$ on its completion provided by observations (e.g. \cite{Cen:2009p1804,Mesinger2009a}).

Another process is proposed in \cite{Dekel:1986p1417}. They argued that the suppression of star formation by supernova feedback in dwarf galaxies embedded in dark matter halos could explain the observed scaling relations in luminosity, metallicity and radius \citep{Dekel:2003p1163}. \cite{Benson:2003p1159} suggested that supernovae could help explain the unexpectedly low dwarf galaxy luminosity function. By removing the gas from galaxies via the injection of thermal and kinetic energy into the interstellar medium, supernovae would reduce the number of stars formed inside dwarf halos. Authors such as \cite{MacLow:1999p1960,Mashchenko:2008p1893,Ricotti:2008p1959,Ceverino:2009p1909} manage to generate massive supernova-driven galactic winds, but \cite{Tassis:2008p1351} claim that including supernova feedback does not affect the properties of their simulated galaxies, attributing the scaling relations found in dwarf galaxies to low star formation efficiencies in weak potentials instead. These contradictions find 
their origin in the different numerical recipes and  numerical resolutions adopted, along with the variety of galaxy masses and merger histories used. As a consequence, it is still unclear as to precisely what effect supernovae have on the ISM gas, and hence on the star formation in low-mass galaxies. Indeed, supernovae are potentially able to drive either positive or negative feedback cycles. Outflows from supernovae can remove gas from the galaxy, preventing it from forming stars. However, they also release metals into the surrounding ISM; metal line cooling increases the efficiency of gas cooling and hence promotes the collapse of gas clouds into star-forming regions \citep{Powell2011}. Moreover, blast wave compression has been observed to trigger star-forming regions \citep{AssousaG.E.1980}.

It has also been recently (re-)suggested that our model of dark matter is incorrect. \cite{Boylan-Kolchin2011,Lovell2012,DiCintio2011} examine the relationship between the maximum circular velocity of the dwarf spheroidal satellites and the radius at which this velocity is found, and determine that the velocities found are higher than those found in either $\Lambda$CDM pure dark matter simulations or simulations with baryon physics using smoothed particle hydrodynamics (SPH). In fact, \cite{DiCintio2011} determine that gas cooling makes the problem worse, since the central density of the halo increases and the radius at which the maximum circular velocity is found decreases. \cite{Lovell2012} suggests that warm dark matter (WDM) resolves the problem, but they do not run simulations with both WDM and baryon physics.

Bearing these caveats in mind, the present paper aims to constrain satellite galaxy formation and evolution, and more specifically the role played by supernova feedback and reionisation in the process. To this end, we use the \Simname ({\sc nut}) suite of high resolution hydrodynamic cosmological simulations of a Milky Way-like galaxy \Nutref. At scales of 1-10pc, these resolve large molecular clouds, and hence model the interstellar gas and stellar feedback in greater detail, which potentially affects star formation histories \citep{Slyz:2005p1032}. This approach differs from previous work investigating the Milky Way satellites using hydrodynamic simulations as these generally capture the ISM at lower resolution, and attempt to compensate for this by introducing analytic expressions to account for the multiphase ISM and outflows  \citep{Scannapieco:2005p1834,Scannapieco:2006p1835,Murante:2010p1836}. As it is currently too costly to simulate a Milky Way-sized halo and its substructures with parsec resolution 
and hydrodynamics throughout the lifetime of the Universe, most of our analysis is restricted to high redshift ($z>6$). However, observational studies conclude that a vast majority of satellite galaxies contain stars which formed prior to $z \sim 2$ and in many cases prior to even $z \sim 5$ (see recent review by \cite{Tolstoy2009} and also \cite{Kirby2011a}). Therefore, a high redshift study of these objects should be able to shed light on the problem, provided one is able to accurately predict their spatial distribution at  $z=0$.

This latter requirement is in itself a major challenge as it presumably requires hydrodynamics simulations which include (at least) radiative cooling. Indeed, since gas cooling can significantly increase the central density of dark matter halos (e.g. \cite{Blumenthal1984}), one expects physical processes like dynamical friction and tidal disruption of the satellites to be altered. The extent of these differences needs to be quantified because most of the studies mentioned earlier in this introduction rely on pure dark matter simulations to underpin analytic arguments or graft semi-analytic models of galaxy formation. Several groups have looked at  differences between simulations of galactic halos containing baryons and their pure dark matter N-body counterparts. For instance \cite{Peirani:2010p2074} found that identical simulations of a local group-like volume with and without baryons matched well, but did not comment on satellites within halos. In their constrained simulations of the Local Group, \cite{
Libeskind:2010p1751} find more satellite halos when baryons are included than in the identical pure dark matter run.Their radial distribution is also significantly more concentrated. By contrast, although they also follow a more concentrated radial distribution, satellites in the baryonic simulations of  \cite{RomanoDiaz:2010p2098,RomanoDiaz:2009p1742} survive for shorter times than their pure dark matter counterparts, which yields an overall lower number of satellites within $R_{\rm vir}$ when baryons are included. \cite{Schewtschenko:2011p2097} finds a similar result to \cite{Libeskind:2010p1751} in terms of number of halos but suggests that these results are not incompatible with those of \cite{RomanoDiaz:2010p2098}. Further, \cite{D'Onghia2010} find that the presence of a disk can affect the mass function of satellites around a host halo. We re-examine this issue using Eulerian AMR grid hydrodynamics instead of Lagrangian smoothed particle hydrodynamics (SPH), and improving on both mass and force 
resolution for the dark matter. 

This paper is split into three main parts. In section \ref{methods}, we describe the simulations, our algorithms for comparing them and for tracking halos at high redshift down to $z=0$. Section \ref{feedback} looks at the effect of feedback mechanisms on satellite galaxy formation at high redshift and ultra-high resolution. The third part, section \ref{satstoday}, is devoted to the present epoch, and how our results at high redshift affect the satellite population we see today. 

\section{Methods}
\label{methods}

In this section we discuss the methods employed to carry out the simulations used in this paper and the subsequent analysis techniques.

\subsection{Numerical simulations}
\label{methods_numsim}

\begin{table*}
\begin{tabular}{c c c c c c c c c c}
   \textbf{Simulation} & \textbf{z$_{min}$} & \textbf{m$_{\rm DM}$} & \textbf{R$_{max}$ (level)} & \textbf{Gas Cooling} & \textbf{m$_{\star}$} & \textbf{SNe} & \textbf{UV}\\
  \hline
 Reference Run & 0 & 5.6$\times10^{4}$\Msolar & 50pc (18) & \tick & 3.5$\times10^4$\Msolar & &\tick\\
 Cooling Run & 6.7 & 5.6$\times10^{4}$\Msolar & 0.5pc (25) & \tick & 167\Msolar & & \tick\\
 Feedback Run & 6.7 & 5.6$\times10^{4}$\Msolar & 0.5pc (25) & \tick & 76\Msolar & \tick &\tick\\
   \hline
 Dark Matter Run & 0 & 6.7$\times10^{4}$\Msolar & 50pc (18) & & - & & \\
 Adiabatic Run & 0 & 5.6$\times10^{4}$\Msolar & 50pc (18) & & - & & \tick\\
\end{tabular}
  \caption{Table of properties of numerical simulations included in this paper. The columns are, from left to right, the simulation name, the lowest redshift reached by the simulation, the minimum dark matter particle mass, the maximum spatial resolution of the AMR grid (with the associated level of refinement in brackets), whether the simulation includes gas cooling, the minimum star particle mass, whether the simulation includes supernova feedback, and whether the simulation includes a UV background. The bottom two simulations are considered only in section For a complete description of the simulations, see section \ref{methods_numsim}.}
\label{methods_numsimtable} 
\end{table*}

We analyse five simulations in the \Simname ({\sc nut}) suite of simulations (\Simname is the Ancient Egyptian goddess of the sky) \Nutref. \Simname is a set of cosmological hydrodynamic resimulations of a Milky Way-like halo at $z=0$ (throughout this paper, this halo will be referred to as the ``Milky Way''). To run these simulations, we use the Adaptive Mesh Refinement (AMR) code {\sc ramses} \citep{Teyssier:2002p533}. Each simulation starts from identical initial conditions, which are generated with {\sc mpgrafic} \citep{Bertschinger:2001p938,Prunet:2008p970} using cosmological parameters consistent with the WMAP 5 year measurements \citep{Dunkley:2009p972}. The simulation volume is a periodic, cubic box of length 9h$^{-1}$Mpc with a minimum resolution of 128$^3$ dark matter particles and the same number of grid cells. Within this volume we carve out a spherical region of radius 1.44h$^{-1}$Mpc, centred on a halo that reaches a virial mass M$_{\rm vir} = 5\times10^{11}$\Msolar at z=0. We place three 
nested grids in this spherical region with effective resolutions of 256$^3$, 512$^3$ and 1024$^3$ dark matter particles and grid cells. The minimum dark matter particle mass inside this region is equal to 5.6$\times10^{4}$\Msolar (with the exception of the Dark Matter run, which we describe later in this section). We then allow the grid inside the refinement region to adaptively refine up to a given maximum level for each simulation. Our refinement strategy is quasi-Lagrangian:  a grid cell is refined when the number of dark matter particles in the cell exceeds 8, or the baryonic mass of the cell reaches 8 m$_{\rm SPH}$, where m$_{\rm SPH} = 9.4\times10^3$\Msolarnsp. The simulation parameters used are summarised in table \ref{methods_numsimtable}, and in the text below.

The three main simulations that we consider in this paper contain dark matter, gas cooling and a uniform UV background switched on at $z = 8.5$ to model reionization \cite{Haardt:1996p1457}. Gas cooling 
is modelled as radiative energy loss from atomic processes including emission line cooling (below $10^4$K), with a primordial metallicity of 0.001 \Zsolarnsp. Star formation in the simulation proceeds according to a Schmidt law on a local dynamical timescale \citep{Cen:1992p1348} with an efficiency of 1\%. The density threshold for star formation is set in each simulation to be comparable to the corresponding Jeans density $\rho_{J}$ of a cell on the highest level of refinement with a temperature of 100 K. $\rho_{J}$ is given by $(\pi c_{s}^{2}) / (\lambda_{J}^{2} G) = k_{B} T / (m_{H} \lambda_{J}^2 G)$ for an ideal gas, where $\lambda_{J}$ is set to the cell length \citep{Binney:2008p1799}.

We first run a simulation that we call the ``Reference run''. The Reference run is allowed to refine adaptively to up to 8 times inside the fixed refinement region, such that the densest regions are allowed to reach a maximum physical resolution of 50pc at all times,  between a few times and an order of magnitude higher than other cosmological hydrodynamics simulations of Milky Way satellites \citep{Libeskind:2010p1751,Nickerson:2011p2094,Okamoto:2009p1754,Parry2012,Ricotti:2005p2099,RomanoDiaz:2010p2098,RomanoDiaz:2009p1742,Sawala2012a,Scannapieco2011a,Schewtschenko:2011p2097,Wadepuhl2010a}. The minimum star particle mass in this simulation is 3.5$\times10^4$\Msolarnsp. We run this simulation to $z=0$. Note that the main purpose of this run is to act as a lower spatial resolution ``twin'' of the two ``high resolution'' simulations in this study, allowing us to determine which of the galaxies formed at high redshift are progenitors of Milky Way satellite galaxies today. For this reason the DM mass resolution 
is kept identical in all runs. The threshold for star formation in the Reference run is 10 atoms/cm$^3$.

We then run two high resolution simulations which are allowed to refine adaptively by up to 15 times so that its physical spatial resolution in the densest regions can reach a maximum of 0.5pc at all times. The first of these high resolution simulations we call the ``Cooling run''. As with the Reference run, the Cooling run contains dark matter, gas cooling and a uniform UV background switched on at $z=8.5$, but now the threshold density for star formation is $10^5$atoms/cm$^3$. As a result, the minimum star particle mass formed in the Cooling run is 167\Msolarnsp. 
The second of the high resolution simulations is called the ``Feedback'' run. The Feedback run is identical to the Cooling run, except that it also includes supernova feedback. Following \cite{Dubois2008a}, supernovae are implemented as Sedov blast waves with a radius of 2 grid cells (1pc) around a star particle 10 Myr after it formed. Note that while we do not resolve individual stars, at this mass resolution and assuming a Salpeter initial mass function \citep{Salpeter:1955p2059}, we get one supernova per star particle. We assume supernova events entrain 50\% of the initial mass of the star particle ($\eta_{\rm W} = 1$ in the notation 
of \cite{Dubois2008a}) in a wind and have a metal yield of 0.1. The energy released is given by $\eta_{\rm SN}\frac{{\rm m}_\star}{{\rm m_{SN}}}{\rm e_{SN}}$, where ${\rm m_\star}$ is the mass of the star particle, m$_{\rm SN}$ and e$_{\rm SN}$ are typical values for a massive star going supernova and $\eta_{\rm SN}$ is the fraction of the total mass formed that is turned into supernova ejecta. For this simulation, we use $\eta_{\rm SN}$ = 0.106, m$_{\rm SN}$ = 10\Msolar and e$_{\rm SN}$ = $10^{51}$ergs \Nutref. This translates into a minimum star particle final mass of 76\Msolarnsp  ~for the Feedback run. 

 The density threshold for the Reference run is chosen to best match the star formation rate per halo measured in the Cooling and Feedback runs, since these capture the length scales of molecular clouds, allowing for a more realistic model of star formation. The density threshold has to obey two constraints: (i) the star formation density threshold should be smaller than the corresponding $\rho_{J} (\sim $ 40 at/cm$^3)$ on the highest level of refinement and (ii) stars should not form in smooth filaments, which yields a lower bound on the density threshold that we empirically determine to be $\sim$ 10 at/cm$^3$ \citep{Powell2011}. A higher threshold will thus limit star formation to cells with a higher average density. Note, however, that this is the average density, and a volume of the ISM of length 50pc with a low average density may still host small regions of high density gas. Hence a gas cell in the Reference Run with a density below $\rho_{J}$ may still form stars. We determine that a value of 10 
atom/cm$^3$ better matches the star formation rates found in the Cooling and Feedback runs. This value was found by using a number of test runs of the Reference Run using different star formation density thresholds.

Finally, we perform a further two simulations. These are called the Dark Matter run and the Adiabatic run. Both have the same initial conditions and refinement criteria as the Reference run. The Dark Matter run is a pure N-body dark matter simulation, in which the mass in baryons is replaced by mass in dark matter, such that a dark matter particle is $1/(1-f_b)$ times the mass of a dark matter particle in the runs containing baryons, where $f_b$ is the universal baryon fraction (0.173, based on the data in \cite{Dunkley:2009p972}). This gives it a minimum dark matter particle mass of 6.7$\times10^{4}$\Msolar rather than 5.6$\times10^{4}$\Msolar, which is the value common to all the other runs. The Adiabatic run is identical to the Reference run, except that the gas is not permitted to radiate away its energy. As a result, no star formation takes place in the Adiabatic run, though we still include the UV background for sake of comparison. These two simulations are used to determine the effect of 
including more physics on satellite galaxy evolution from $z \sim 6$ (the redshift where 
the high resolution simulations stop) to $z = 0$.
 We discuss the results of this study in section \ref{satstoday_physics} and compare the Dark Matter run to other pure N-body dark matter simulations of Milky Way satellites in section \ref{satstoday_dm}.

\subsection{Halo identification}
\label{methods_halomaker}

We use HaloMaker to identify dark matter halos and galaxies in each simulation output using the Most Massive Subhalo Method (MSM) \citep{Tweed2009a}. We define independent halos as dark matter overdensities not contained within another halo, and subhalos as halos that are identified as substructures of other halos. Similarly, we define galaxies as overdensities in the star particles. An overdensity is defined as a structure which is above $178\rho_{crit}$, where $\rho_{crit}$ is the critical density of the universe. The method works as follows. The density of each particle is found using the SPH technique (e.g. \cite{Springel2001}). Peaks in the density field are then identified, which correspond to the centre of a halo or galaxy. Particles are then attached to a given peak by stepping through decreasing density thresholds, and assigning each particle above this threshold to the nearest peak. Saddle points in the density field are identified, which are used to construct a tree of peaks, 
truncated at $178\rho_{crit}$. Each leaf of this tree is a halo or subhalo. The host halo is identified as the most massive peak, while the other peaks become subhalos. For this procedure, we reject any identified halo that contains less than 40 particles, twice the absolute minimum threshold before spurious halos are identified given in \citep{Tweed2009a}). We also reject any galaxy that contains less than 10 particles -- this figure is lower because stars are typically more tightly clustered than dark matter halos). The minimum total mass of a given dark matter (sub)halo is thus 2.2$\times10^{6}$\Msolarnsp and the minimum stellar mass for a galaxy is 3.5$\times10^5$\Msolar in the Reference run, 1700\Msolar in the Cooling run and 760\Msolar in the Feedback run. 

We perform this process in every simulation for both the dark matter and the stars whenever possible. We thus identify every dark matter halo and every galaxy above the mass limits given in the last paragraph. We visually inspect the results of the halo identification by overplotting the halos on a projection of the density field and tune the halofinder parameters such that we minimise spurious halo detection or halos that are visually identifiable but not detected by the halo finder. 

We then identify the host halo of galaxies to determine which halos are luminous and which are dark. This sorting proceeds in two steps: (i) for each galaxy, we make a list of halos that enclose it within their virial radius (ii) we select the halo that lies closest to the galaxy centre. This final step is needed when the galaxy's host is a subhalo of a larger halo, and the galaxy lies within the virial radius of both the subhalo and its host.
This allows us to match halos between runs and to compare the properties of the embedded galaxies in each simulation on an individual halo basis.

\subsection{Halo twinning}
\label{methods_twinning}

\begin{figure*}
\centerline{\includegraphics[width=1.0\hsize]{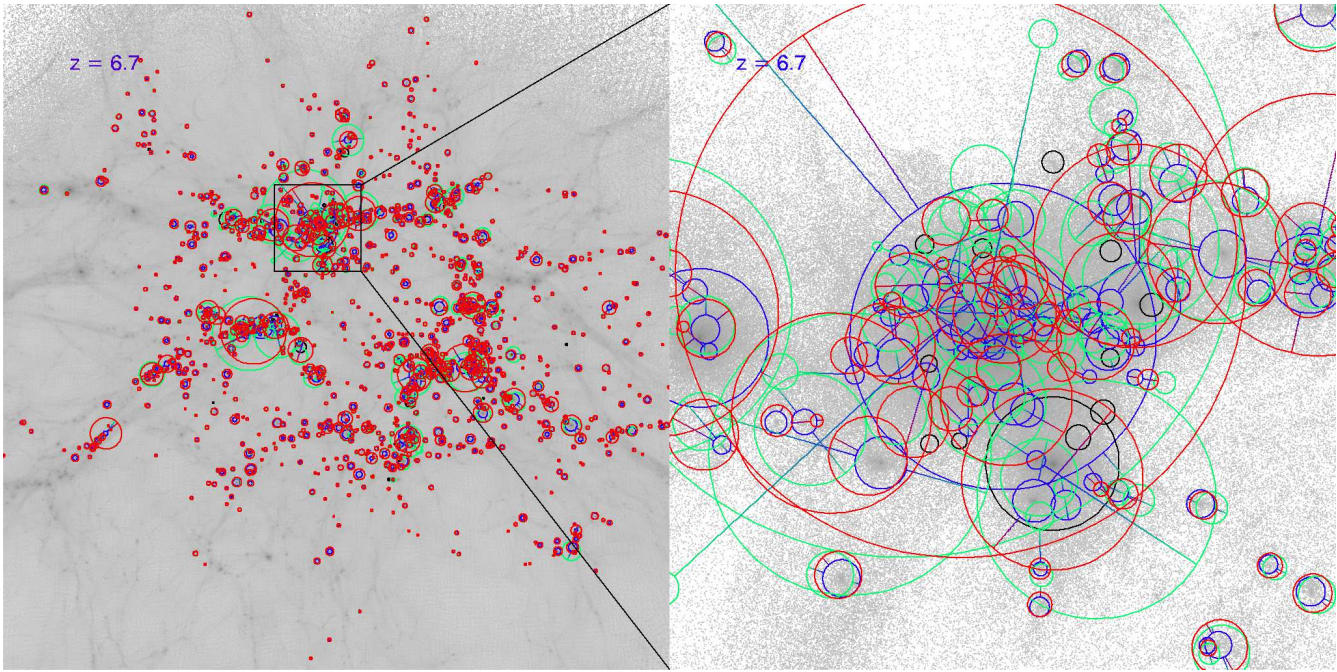}}
  \caption{Visualisation of the halo twinning results at $z=6.7$, showing all successfully twinned halos. The left panel shows the cubic volume (300 physical kpc on a side) containing all the Milky Way satellite progenitor dark matter halos identified by $z=0$. The right panel shows a zoom on the cubic volume containing the Milky Way progenitor halo outlined by the black square in the left image (46 physical kpc on a side). The grey scale background represents the dark matter projected density distribution in the Reference run. Overlaid circles indicate the virial radii of halos identified as being Milky Way progenitors in the Reference run (blue) and their twins in the Cooling (green) and Feedback (red) runs, with colours overplotted in that order (hence halos with very similar positions and radii in all three runs appear as red circles). A black circle is a halo in the Reference run that has no identifiable twin in the Cooling or Feedback run. In the right-hand image we connect the 
halo in the Reference run with its twin in the other runs via straight lines. The Milky Way progenitor halo in each run is shown in black. Most of the twins are remarkably well matched in size 
and position, although unsurprisingly, the subhalos of the Milky Way progenitor show more pronounced discrepancies between runs, especially in the central region of the halo.}
\label{dm_overview}
\end{figure*}

\begin{figure}
\centerline{\includegraphics[width=1.0\hsize]{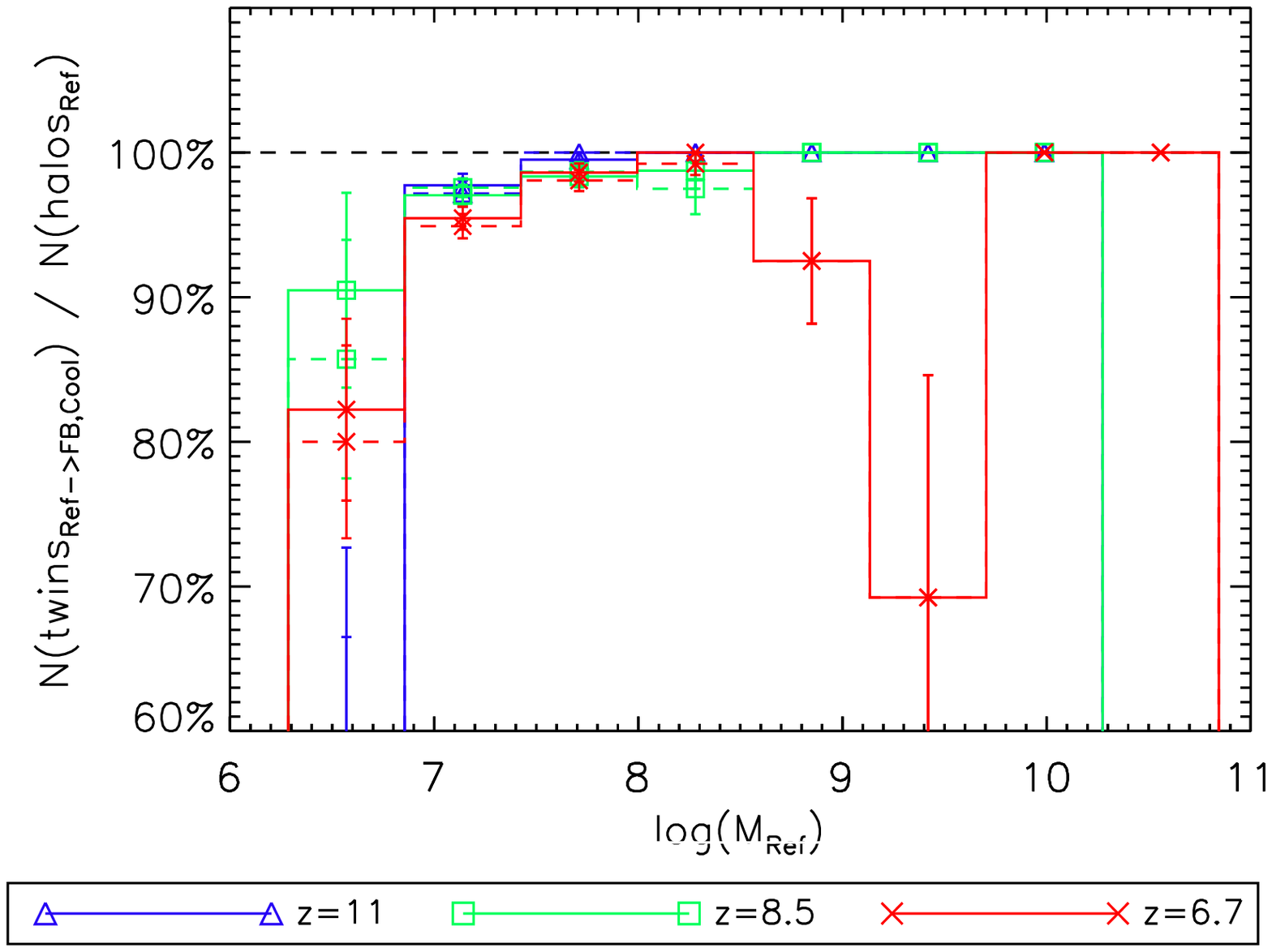}}
 \caption{Histograms of percentage of halos in each mass bin in the Reference run with twins in the Cooling and Feedback run against virial mass in the Reference  run in \Msolarnsp. The error bars show the sampling error on the number of un-twinned halos in each bin. The colours correspond to the values at different redshifts (see legend). Twins in the Feedback Run are shown with a solid line, and twins in the Cooling Run are shown with a dashed line (for mass bins above $10^{8.6}$ the Cooling Run and Feedback run data are identical and hence the lines overlap). For halos of mass greater than $10^7$\Msolarnsp, we find a success rate above 95\% in the twinning procedure in most mass bins. For halos below $10^7$\Msolar (less than 200 DM particles) and at higher redshifts, this rate drops quite rapidly because of the lower spatial (force) resolution in the Reference run which reduces the overall number of collapsed objects. For the mass bins $10^{8.5}$--$10^{9.7}$\Msolar, merger activity 
amongst the relatively low number of halos in these mass bins causes a drop in the success rate of the twinning at z=6.7 to 93\% and 70\% respectively.}
\label{graph_of_twins}
\end{figure}

In order to compare the simulations, we associate each halo in the Reference run with a counterpart (called `twin') in the Cooling and Feedback runs. 
As well as allowing us to compare halos between different runs at common redshifts, this procedure also permits us to track the halos in the Cooling and Feedback runs down to $z=0$ via the 
merger trees of their counterparts in the Reference run (see section \ref{methods_tracking}).

We adopt a twinning strategy similar to \cite{Libeskind:2010p1751,Peirani:2010p2074,RomanoDiaz:2010p2098,RomanoDiaz:2009p1742,Schewtschenko:2011p2097}. A list of particles and associated halo ID numbers is found for each halo in each simulation for each output in the Reference run. Due to the differences in the timestepping in each simulation there is typically 0.5-1Myr difference between a given Reference run output and a given output in the Cooling and Feedback runs. The list of halos in the Reference run is sorted in descending order of halo mass. Particles in the Reference run list are removed if they are not in halos in the Cooling or Feedback run. If a certain halo in the Reference run has fewer than 50\% of its particles in halos in the Cooling or Feedback run, it is considered to have no identifiable counterpart in the other simulations and is hence ignored.

For each halo in the Reference run we calculate the fraction of its particles that belong to a halo in the Cooling and Feedback runs. We then select the single halo from each of the Cooling and Feedback runs that has the largest fraction of its particles in the Reference run halo and has not already been assigned to another Reference run halo. This provides a 1:1 mapping between the halos of any two simulations.

In order to visually confirm that the twinning procedure works, we plot a map of the Reference run's projected dark matter density field in Fig.~\ref{dm_overview}. On the same figure, we overplot halos in the Reference, Cooling and Feedback runs as circles of radius r\Subvir. We also link the halos in the Reference run to their twins in the Cooling and Feedback runs with straight lines connecting the 
corresponding circles. The figure clearly shows that in general, the twinning procedure yields excellent results for most halos (the vast majority of circles in the left panel of Fig~\ref{dm_overview} are red).
For the region encompassed by the Milky Way progenitor (the right-hand zoomed-in panel), twinning results are still quite good, except in the very centre where 
 positions and sizes of sub-halos diverge as the non-linear nature of the system (shell-crossing) and the slight differences in output times between the runs begin to plague
 the comparison.

A quantitative analysis of the twinning procedure reveals that the Reference run has 96.4\% of its halos twinned with the halos in either the Feedback or Cooling run at z=6.7 (the final redshift for which all runs have data). If we relax the 1:1 mapping criterion and simply consider halos above the threshold where 50\% of their particles in the Reference run also are found in halos in the other runs, 98.6\% of halos have twins. In Fig. \ref{graph_of_twins}, we plot the success rate for twinning halos as a function of mass and redshift for the Feedback (solid lines) and Cooling runs (dashed lines). We find that for halos over 10$^9$\Msolar there is a 100\% success rate for the twinning procedure at z=8.5 and above. At z=6.7, we find that the mass bins $10^{8.5}$--$10^{9.7}$\Msolar exhibit a drop in twinning success rates to z=6.7 to 93\% and 70\% respectively. This is due to the non-linearity of the N-body problem and merger activity as described above, as well as the relatively low number of halos in 
these mass bins. The twinning success rate is over 95\% above 10$^7$\Msolar (i.e. for halos containing $\gtrsim$ 200 DM particles). For halos below this mass, the lower resolution of the Reference run causes the success rate of the twinning procedure to drop to between 70-90\%. Note that only the {\em spatial} (or 
force) resolution in the Reference run is lower than in the Cooling or Feedback runs; the dark matter mass resolution is identical.

\subsection{Tracking high redshift galaxies down to $z=0$}
\label{methods_tracking}

The ultimate goal of this project is to compare our simulated galaxies to observed Milky Way satellites. In order to achieve this, we need to evolve our simulated galaxies in the Cooling and Feedback run to $z=0$. Since it is computationally unfeasible at their nominal resolution, we instead track their evolution via their twin halos merger trees in the Reference run. This determines which galaxies at high redshift are the progenitors of Milky Way satellites today and allows us to quantify how advanced satellite galaxy formation is by the end the epoch of reionisation.

The fundamental assumption we make is that a halo which already contains stars at high redshift will still contain a galaxy at $z=0$.
This assumption is extremely plausible for two reasons. Firstly, even the lowest-mass halos are observed to be dark-matter dominated \citep{Strigari:2008p994}, and thus we do not expect to find galaxies without dark matter halos. Secondly, galaxies are all predicted to be embedded within the inner part of the halo in which they form, so that the galaxy will be the last part of the halo to be destroyed, with tidal stripping affecting the outer regions of the halo first (e.g. \cite{Penarrubia:2010p2073}). We further comment on the validity of this assumption in section \ref{satstoday_z0} where we identify galaxies in the Reference run at z=0 and locate their dark matter host halos. Finally, we also assume with this extrapolation technique that the dynamical friction and tidal stripping experienced by the satellite halos in the Reference Run are similar to the Cooling and Feedback runs, i.e. that increased resolution and supernova feedback do not dramatically alter their efficiency. We discuss the validity of 
this assumption in more detail in section \ref{satstoday_physics}.

We build the merger tree for the Reference run using the Branch History Method (BHM). BHM compares the subhalo population of a given host halo between two snapshots, and attempts to optimise the tree structure to account for anomalies such as subhalos without an identified progenitor, or a host and subhalo switching place during a major merger event; for details of this technique, see \cite{Tweed2009a}. As in section \ref{methods_twinning}, we use the particle IDs to track dark matter particles between snapshots. For every halo in a given snapshot we build a list of halos in the following snapshot that contain particles from this halo. We then select the halo that contains the most particles from this halo as its ``child'' halo, adopting a ``one child'' policy. By doing this, we create a halo merger tree where if halo C in output 3 is a child of halo B in output 2 and halo B is a child of halo A in output 1, then halo C is also a child of halo A. One side-effect of this method is that if a subhalo 
loses more than 50\% of its particles between two outputs, that subhalo is assumed to have been completely stripped by its host. To limit this occurrence, we use a large number of snapshots to build our merger tree ($\sim$ 100), so that our effective time resolution is roughly 150Myr.

\subsection{Resolution effects on galaxy formation}
\label{resolution}

\begin{figure}
\centerline{\includegraphics[width=1.0\hsize]{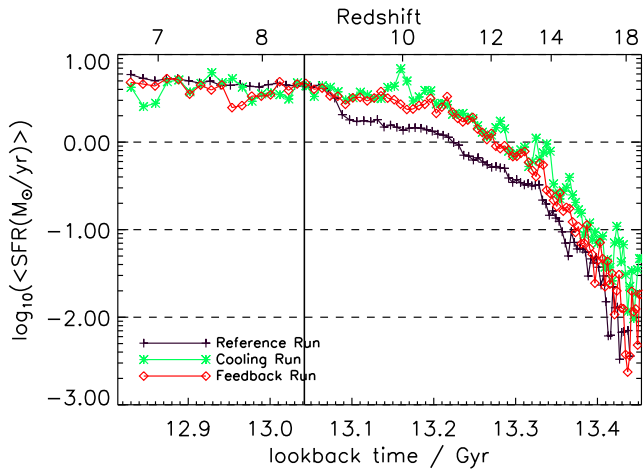}}
  \caption{Star formation rate averaged over all halos in each simulation. The Reference run is show in black, the Cooling run in green and the Feedback run in red. The vertical black line shows the time ($z=8.5$) at which the universe is reionised in the simulation. Star formation in the Reference run is slightly delayed compared to the other runs, but catches up before reionisation. The jump in star formation rate at $z=9$ is due to the triggering of a new level of refinement on the grid (from 14 to 15 levels) at this redshift, which allows the gas to collapse further and triggers star formation in all potentially star-forming halos.}
\label{sfr_compare}
\end{figure}

\begin{figure}
\centerline{\includegraphics[width=1.0\hsize]{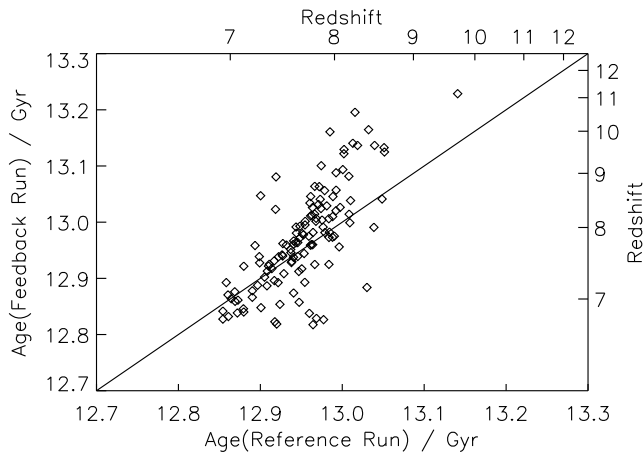}}
  \caption{Comparison of the mass-weighted stellar age for halos in the Reference run and their twins in the Feedback run. Halos that lie on the diagonal black line have the same mass-weighted stellar ages in both runs. Due to its lower resolution, as explained in section \ref{resolution}, the Reference run generally forms stars later than the Feedback run. However, the difference in star formation parameters between the runs allows a modest fraction of halos to form stars before, especially at lower redshifts. Similar results are found when comparing the Reference run to the Cooling run.}
\label{starages_compare}
\end{figure}

In order to assess how reliable the Reference run is to locate satellite galaxies at $z=0$, we first compare it to the higher resolution Cooling and Feedback runs in the redshift range where all the runs overlap.
In section \ref{methods_twinning}, we showed that we are able to very successfully match halos more massive than 10$^7$\Msolar between simulations. 
We now consider the effect that resolution has on star formation.

As previously mentioned in section \ref{methods_numsim}, the Reference run has the same dark matter mass resolution as the Cooling and Feedback runs. However, the spatial resolution, which determines the accuracy of both the gravitational force and the properties of the gas is lower; 50pc in the Reference run instead of 0.5pc in the Cooling and Feedback runs. The density threshold for star formation is therefore lowered
from $10^5$atoms/cm$^3$ at high resolution down to 10atoms/cm$^3$ at low resolution, whilst the efficiency of star formation is preserved. 

In figure \ref{sfr_compare}, we compare the global star formation rate of halos in each of the runs. We find that before a lookback time of 13.1Gyr ($z=9$), the Reference run's star formation rate is roughly half that of the Cooling and Feedback runs. However, after 13.1Gyr, all star formation rates agree within 30\% percent. This difference of behaviour before and after 13.1Gyr has nothing to do with reionisation, which occurs later on. Indeed this effect is purely numerical, and induced by the refinement criteria we choose to enforce. {\sc ramses} refines the AMR grid using an octree, meaning that spatial resolution is a power-of-two fraction of the total box length \citep{Teyssier:2002p533}. Since we specify a maximum spatial resolution for the grid in physical parsecs, we trigger a power-of-two increase in resolution each time the cosmological scale factor has increased enough that an extra level is necessary to achieve such a resolution. In the Reference run, such a jump in refinement level from 14 to 
15 happens around $z=9$. \cite{Rasera:2006p2060} demonstrate that too low a maximum spatial resolution delays the collapse of low mass haloes/galaxy disks, preventing the ISM gas density in many of them from crossing the star formation threshold until a higher level of resolution is achieved. Such a delay eventually vanishes when the maximum spatial resolution becomes sufficient {\em at all times} as the lack of `step' in the star formation histories of the Cooling and Feedback runs on Fig~\ref{sfr_compare} clearly shows. After a lookback time of 13.1Gyr, the star formation rates in all runs match well, since as discussed in section \ref{methods_numsim}, we select a density threshold for star formation in the Reference Run that best matches the star formation rates in the higher resolution simulations.

We now consider the agreement between star formation histories of individual galaxies. In Fig. \ref{starages_compare}, we compare the mass weighted stellar ages of galaxies simulated at low (Reference run) and high (Feedback run) resolution and twinned at $z=6.7$.  We find a pattern similar to that of the global star formation histories presented in Fig. \ref{sfr_compare}; namely star formation is delayed in the Reference run but converges to values similar to the Feedback run at a lookback time comprised between 13.0 and 12.9 Gyr. After this epoch there is some inevitable scatter due to the nonlinear nature of star formation, but this scatter is centred around the line of equal age in Fig. \ref{starages_compare}. The lookback time at which the ages converge is later than the jump in star formation due to resolution because the mean age is skewed by the relative paucity of stars formed before 13.1 Gyr in the Reference run (Fig. \ref{sfr_compare}). A similar result is found when comparing the Cooling run to 
the Reference run.

It is worth pointing out that the location of star formation within a galaxy is not guaranteed to match between simulations. Star formation in the Cooling and Feedback runs is confined to regions with densities similar to molecular cloud cores ($\rho > 10^5 $atoms/cm$^3$), whereas in the Reference run star formation is allowed to occur in regions where the density is closer to that of typical diffuse clouds (10 at/cm$^3$). However, for analysing the bulk properties of satellite galaxies between reionisation and $z=0$ this distinction is largely irrelevant.

\section{Feedback in Milky Way satellite progenitors}
\label{feedback}

\subsection{Supernova Feedback}
\label{feedback_sne}

\begin{figure*}
\centerline{\includegraphics[width=1.0\hsize]{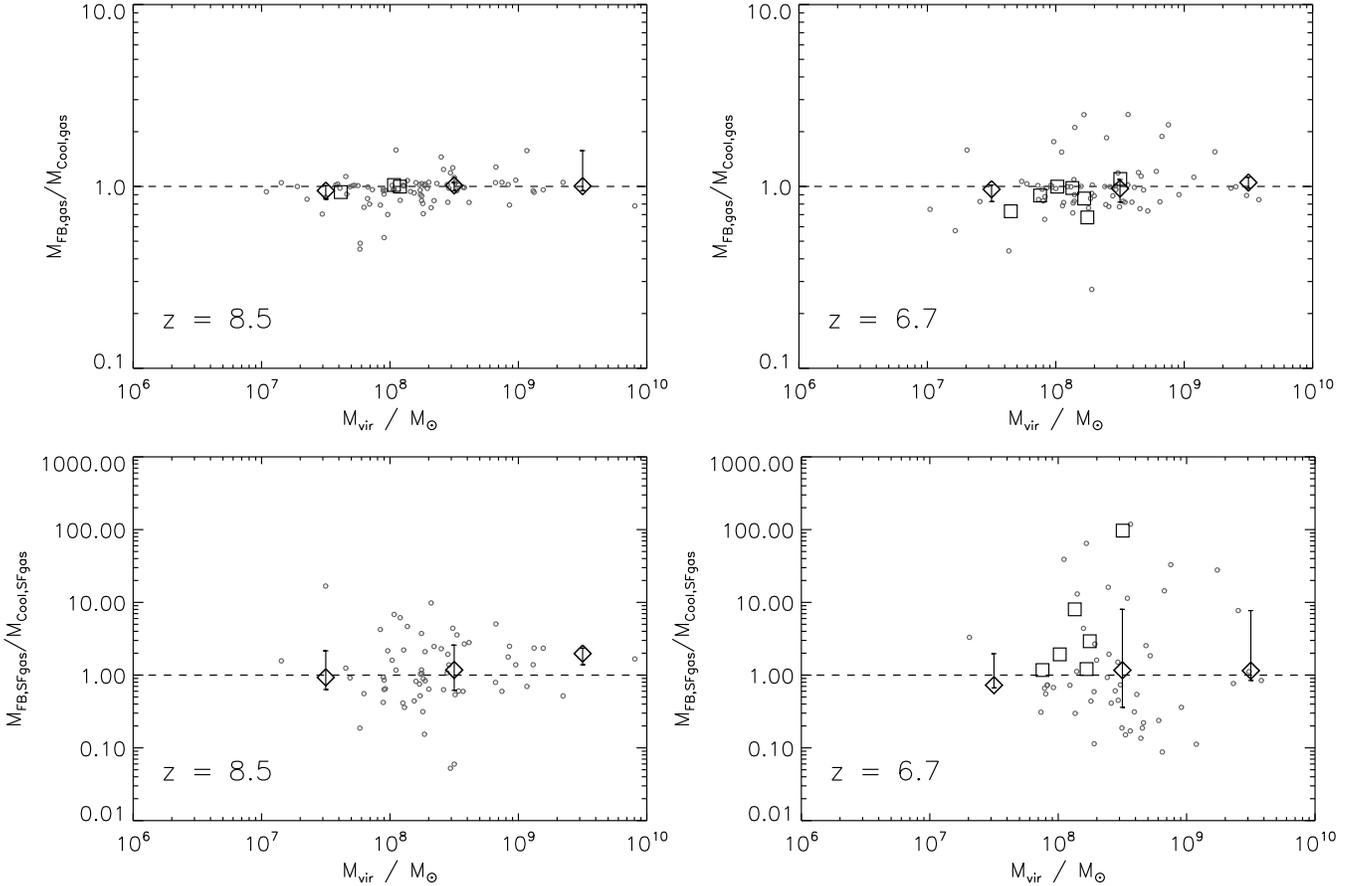}}
  \caption{Ratio of gas mass in galaxies within r\Subvir in the Cooling and Feedback runs, plotted against the total halo mass of their twin halo in the Reference run. The figure shows galaxies at the redshift of reionization $z=8.5$ on the left , and $z=6.7$ on the right. The top plots show the ratio of total gas mass between runs, while the bottom plots show the ratio of the star-forming gas mass (i.e. $\rho > 10^5$atoms/cm$^3$). A black square indicates that the halo of that galaxy survives as a Milky Way satellite at $z=0$; grey circles are halos that are completely disrupted by $z=0$. We overplot as diamonds with error bars the median and interquartile range of the fractional differences in halo mass bins $10^7$\Msolar -- $10^8$\Msolar, $10^8$\Msolar -- $10^9$\Msolar and $10^9$\Msolar -- $10^{10}$\Msolar. The median values lie around the horizontal line marking an equal ratio, with the $10^7$\Msolar to 
$10^8$\Msolar bin having a ratio of $\sim 0.95$ and the higher mass bins having a 1:1 ratio or higher. There is a large amount of scatter in the relative amounts of star forming gas in halos in the two simulations. It is worth noting that some of the halos at each redshift sampled, including all of the Milky Way satellite progenitors at $z=8.5$, do not contain star-forming gas. This is explained as star formation occurring in bursts, with the smaller galaxies containing no star-forming gas at certain instants in time.}
\label{totalgascompare}
\end{figure*}

\begin{figure*}
\centerline{\includegraphics[width=1.0\hsize]{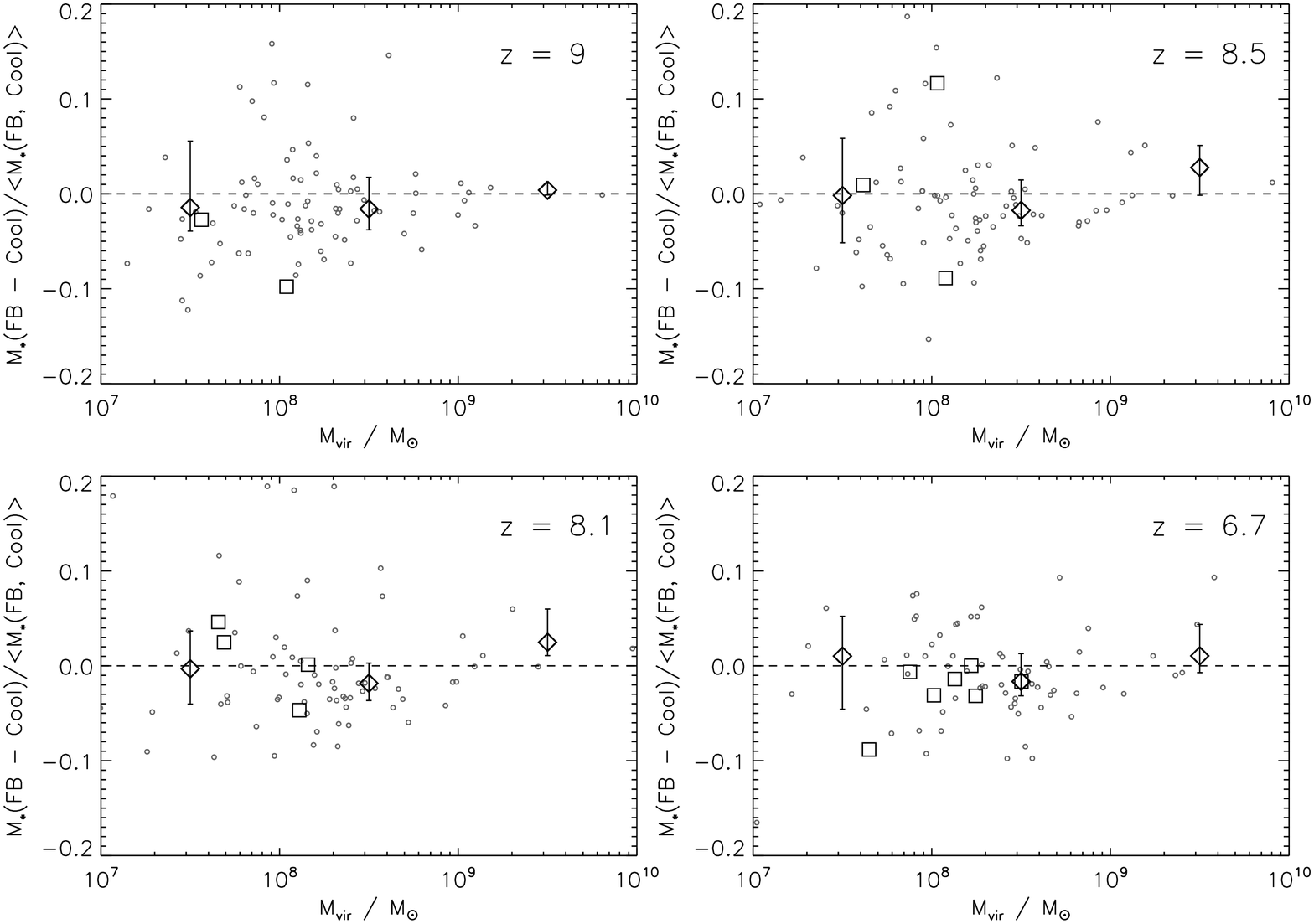}}
 \caption{Fractional difference between stellar masses of galaxies in the Cooling and Feedback run (M$_{\star \rm {FB}}$ - M$_{\star \rm{Cool}}$) / (0.5(M$_{\star \rm{FB}}$ + M$_{\star \rm{Cool}}$)) versus total twin halo mass in Reference Run, where M$_{\star \rm{FB}}$ is the stellar mass in the Feedback run and M$_{\star \rm{Cool}}$ is the stellar mass in the Cooling run). A positive value indicates that the inclusion of supernova feedback enhances star formation in the given galaxy, while a negative value means that supernova feedback suppresses star formation. Different panels show the values for different output redshifts, from $z=9$ (top left) to $z=6.7$ (bottom right). Black squares represent  halos containing galaxies which survive as a Milky Way satellites at $z=0$; a halo represented by a grey circle is completely disrupted by $z=0$. We overplot as diamonds with error bars the median and interquartile range of the fractional differences in halo mass bins $10^7$\Msolar -- $10^8$\Msolar, $10^8$\
Msolar -- $10^9$\Msolar and $10^9$\Msolar -- $10^{10}$\Msolar. Star formation is slightly suppressed in low mass halos with weaker gravitational potentials. The trend is reversed for high mass halos.}
\label{starmasscompare}
\end{figure*}

We now discuss the differences between the simulations with and without supernova feedback, in order to better understand the role of supernovae in high-redshift dwarf galaxy and Milky Way satellite formation. We use the twinning method described in section \ref{methods_twinning} to match halos in the Cooling and Feedback runs. We can thus determine whether the net effect of including supernovae in our sub-parsec resolution simulations enhances or suppresses star formation in halos of various masses. We focus on halos that are captured by the Milky Way by z=0, and hence we can determine to what extent supernova feedback influences the star formation history of Milky Way satellites observed today. Whether these high-redshift galaxies become satellite galaxies of the Milky Way at $z=0$ or are disrupted by interactions with the Milky Way halo is discussed in section \ref{satstoday_tracking}.

In Figs \ref{totalgascompare} and \ref{starmasscompare}, we quantify the extent to which the positive feedback processes (metal cooling, blastwave compression) or negative feedback processes (gas heating, outflows) dominate in halos of different masses. We compare the total gas mass, star-forming gas mass and total stellar mass in each halo in the Cooling and Feedback runs, using the halo twinning procedure described in section \ref{methods_twinning}. Star-forming gas is defined as gas with a density above  $10^5$ at/cm$^3$ our density threshold for star formation. In each figure we overplot the median and interquartile range of the fractional differences in the mass bins $10^7$\Msolar -- $10^8$\Msolarnsp, $10^8$\Msolar -- $10^9$\Msolar and $10^9$\Msolar -- $10^{10}$\Msolarnsp. For the gas masses, since there exists  a large scatter in the results we plot the ratio for each halo on a log scale to highlight both large and small differences. For the stellar masses, since differences are smaller, we plot 
the fractional difference between the runs. In other words,  if we denote the ratio between stellar masses by R, we plot the quantity  $(R-1)/(0.5(R+1))$.

Although Fig \ref{totalgascompare} shows a large scatter in ratios of gas masses of twinned halos in the Cooling and Feedback runs, the median values lie around an equal ratio, with the $10^7$\Msolar to $10^8$\Msolar mass bin showing $\sim$ 5\% less gas and the halos in the higher mass bins having a similar or slightly higher gas content in the Feedback run. This goes in the expected direction since supernovae eject gas back into the galaxy, causing the total gas mass to increase if this ejecta is unable to escape the halo. There is also considerably more scatter in the instantaneous star-forming gas mass results, with some halos in the Feedback run containing over 100 times the mass of star-forming gas than their Cooling run twins. This effect can be attributed to the enhanced metal cooling which takes place after the first supernovae explode in the Feedback run. However, on average we find a similar pattern to the total gas mass ratios, i.e. lower mass halos in the Feedback run contain very slightly less 
star-forming gas than their Cooling run counterparts, and more massive halos slightly more. That said, at $z=8.5$ the highest mass bin has a median star-forming gas mass that is twice as high. The large scatter in the amount of star-forming gas is expected; \cite{Stinson:2007p1875} also find that star formation in their dwarf galaxies is quite bursty, because the instantaneous mass of star-forming gas can strongly fluctuate on short timescales, driven by catastrophic non-linear events (instabilities, mergers). 

Motivated by this result, in Fig. \ref{starmasscompare} we plot a time integrated quantity -- the fractional difference between the stellar mass of each twinned halo that has formed stars in the Cooling and Feedback runs. Unsurprisingly, there is a maximum 20\% difference between values, much smaller than that for the star-forming gas mass.  For similar reasons, we also find more scatter in stellar mass ratio of low mass halos than of high mass halos: the length of time that lower-mass halos have been forming stars is generally shorter. Therefore, they are more affected by temporary fluctuations in their star formation histories. As with the gas mass comparison, we find that the positive feedback processes outweigh the negative feedback processes in the highest mass bin, leading to a net increase in the median fractional difference in stellar mass of a few percent when supernova feedback is added. 

In summary, we find that the effect of feedback on the gas mass and star formation in a halo is complex, with lower mass halos being on average more affected by negative feedback processes such as outflows and gas heating, and higher mass halos by positive feedback processes such as blastwave compression and metal cooling. We also find that stellar masses in individual twin halos can differ by up to 20\%; however, the median values only differ by a few percent in all mass bins, although values for low mass halos are more scattered. We therefore conclude that supernovae do not seem to have a significant effect on the total stellar mass of Milky Way satellite progenitors, at least at redshifts larger than 6. 

\subsection{Reionisation Feedback}
\label{feedback_reion}

\begin{figure*}
\centerline{\includegraphics[width=1.0\hsize]{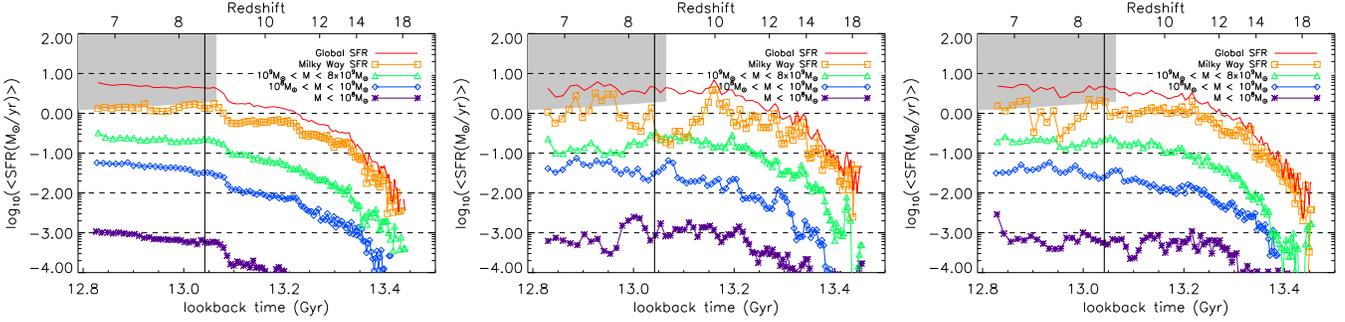}}
  \caption{Mean star formation histories for different halo mass bins in the Reference run (left) Cooling run (middle) and Feedback run (right). Mean star formation rates (SFR) in \Msolar/yr are given for halos of mass M $< 10^8$\Msolar (purple asterices), $10^8$\Msolar $< $M $< 10^9$\Msolar (blue diamonds) and $10^9 $\Msolar$<$M$<8\times10^9 $\Msolar (green triangles), as well as the Milky Way (orange squares) and the global SFR for the entire high resolution region (red solid line). The vertical line at $z=8.5$ (look-back time 13.042Gyr) shows the point at which reionisation is turned on in the simulations. The grey region shows the detectable star formation rates as determined by \protect\cite{Wilkins2011}. This suggests that the SFR for a Milky Way-like galaxy progenitor is almost detectable at z$\sim$8 (lookback time $\sim$13.0Gyr). We find no sudden drop in star formation in any mass bin after reionisation for any of the simulations. In fact, some of the star formation rates increase by up to 0.5 dex 
at reionisation. The jumps in star formation rates in the Reference run are due to the triggering of a new level of refinement on the grid, which allows the gas to collapse to allow star formation in all potentially star-forming regions \protect\citep{Rasera:2006p2060}.}
\label{sfrplots}
\end{figure*}

\begin{figure*}
\centerline{\includegraphics[width=1.0\hsize]{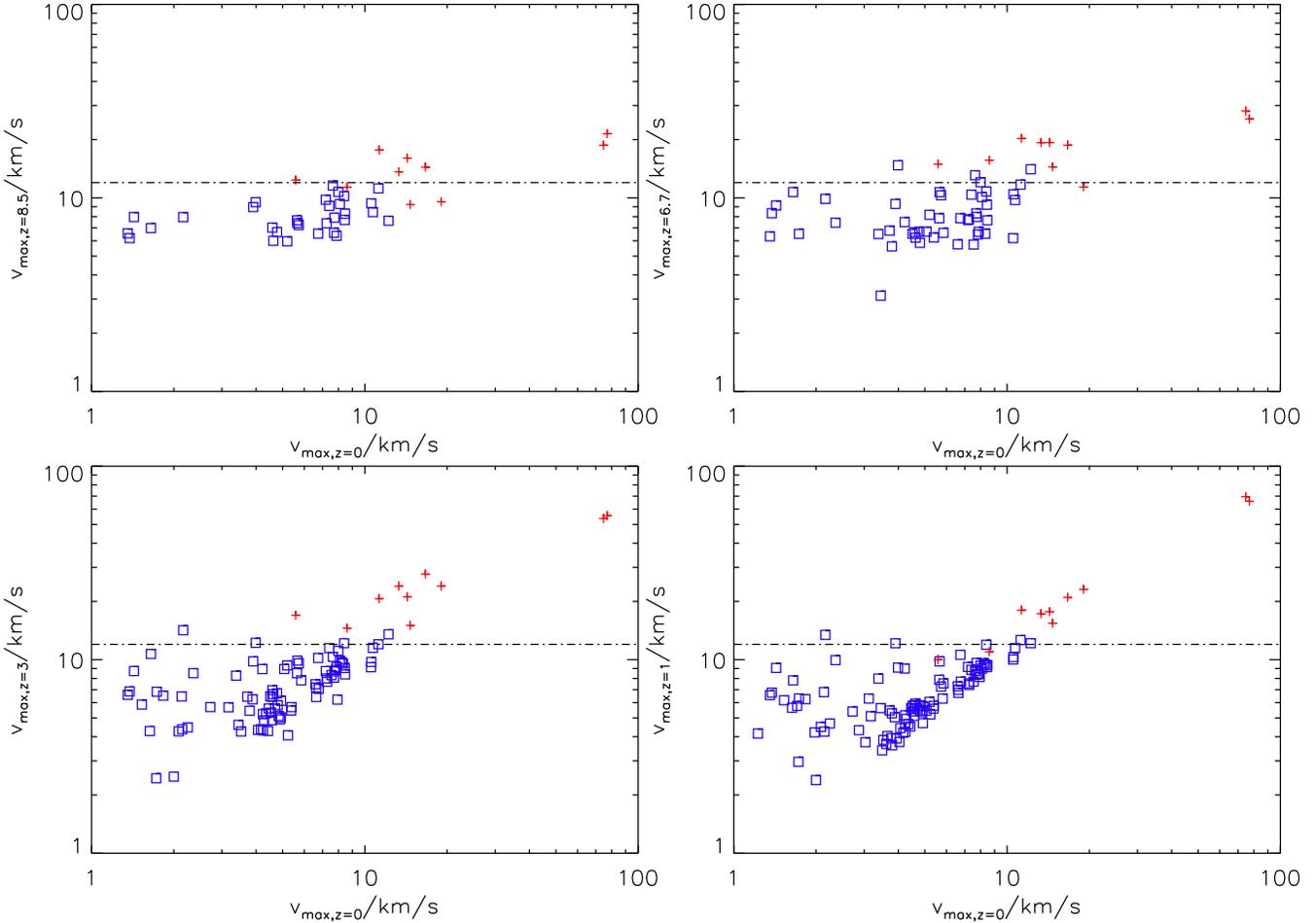}}

  \caption{Maximum circular velocity of satellite halos at $z=0$ versus the maximum circular velocity of their main progenitor at, from top left to bottom right, $z=8.5$ (just prior to reionisation), $z=6.7$, $z=3$, $z=1$. In red are the satellites that contain stars at $z=0$ in the Reference run; the blue satellites remain dark. The dotted horizontal line is at 12km/s and represents the threshold given in \protect\cite{Okamoto:2009p1754} above which halos can form stars before reionisation. Like these authors, we find halos that form stars under this threshold to about 10km/s; below this, no halos can form stars. Note that this threshold seems independent of redshift and that surviving Milky Way satellite galaxies begin to be captured by the Milky Way progenitor at $z=3$. The halos with v$_{max,z=0}$ around 80km/s are satellites labelled `a' and `h' (see Fig. \ref{dm_z9_to_z0}).}
\label{vmax_zhighvsz0}
\end{figure*}

At $z=8.5$ in each simulation we include a simple instantaneous, uniform heating term that represents the UV background according to the model of \cite{Haardt:1996p1457}. There are two ways such a background affects galaxy formation/evolution in our simulations: (i) it heats the ISM, which could prevent the gas density within galaxies from crossing the star formation threshold, and (ii) it heats the IGM, which could cut off the gas accretion onto galaxies. Note that we very likely overestimate these effects in the simulation as we neglect self-shielding which is known to occur around densities $n_H \lesssim 0.1$ at/cm$^3$ \citep{Susa2004}. On the other hand, we do not account for the local UV radiation (from stars within the galaxy itself) which also photo-ionises the ISM/IGM. Nevertheless, by looking at the star formation rate in halos before and after $z=8.5$, we should be able to estimate how efficient non-local ionisation is at halting star formation.

Fig. \ref{sfrplots} shows that reionisation does not immediately stop star formation in halos already forming stars, in agreement with \cite{Kitayama2001,Machacek2001,Gnedin2006,Okamoto:2009p1754,Wadepuhl2010a}. Even for the lowest mass halos (M$_{\rm vir} < 10^8$\Msolar), we find that star formation continues after the uniform UV background is turned on. The free-fall time of a test particle falling from r\Subvir into one of the smallest galaxies formed at $z=8.5$ is on the order of 50 Myr. Hence, we conclude that star formation is not stopped in these halos even after $\sim$5 free-fall times (the amount of time elapsed between $z=8.5$ and $z=6.7$). In other words, if reionisation does halt star formation by heating up the ISM or cutting off the gas accretion in halos that have already formed stars, it does not do so abruptly, but rather over a significantly extended period of time.

Another potential effect of reionisation is to quench galaxy formation by preventing the collapse of gas within halos that have not yet formed stars (e.g. \cite{Gnedin:2000p1832,Somerville:2002p1760,Benson:2002p1802}). However, the last $z=0$ satellite galaxy to be formed in our Reference run begins forming stars at $z=4.8$ in a halo with M$_{\rm vir} = 1.4\times10^7$\Msolar, and there are 9 other satellites galaxies hosted by halos with a similar mass which form their first star after $z=8.5$. Hence, whilst it is still possible that UV photoionisation has a long-term role in preventing some galaxies from forming, it does not seem to be able to halt galaxy formation entirely. 

In figure \ref{vmax_zhighvsz0}, we recast this statement in terms of minimal circular velocity for a star forming halo, v$_{max}$, below which halos are prevented from forming stars. This allows us to directly compare our results to those presented by \cite{Okamoto:2009p1754}. As these authors, we cannot definitively conclude that this threshold arises entirely because of reionisation or the general inability of halos below a v$_{max}$ of 10km/s to cool and form stars by $z=0$, since we do not run a simulation without reionisation. However, we note that reioinisation in our simulation occurs instantaneously at $z=8.5$ (close to the value of $z=9$ of \cite{Okamoto:2009p1754}) and that, in stark contrast the threshold of v$_{max} \approx$ 10 km/s seems independent of redshift. Indeed, it remains quite constant both before and after reionisation has occurred, which leads us to argue that reionisation cannot play an important role in setting its level and only sustains it, in the best of cases. 

\section{Milky Way satellites today}
\label{satstoday}

\subsection{Tracking satellite galaxies down to $z=0$}
\label{satstoday_tracking}

In section \ref{methods_tracking}, we discussed the techniques used to track the galaxies formed in the Cooling and Feedback run down to $z=0$ using the Reference run. We now analyse the results of this tracking. We resolve about 6630 such halos. Of these 6630, 394 survive as subhalos of the Milky Way halo at $z=0$. 
In table \ref{star_fate}, we list the number of galaxies formed by various redshifts from $z=11$ to $z=6.7$ in the Cooling and Feedback runs that survive to $z=0$. It is apparent from the table that by this redshift, we have not formed enough satellite galaxies to match even the population of pre-SDSS satellites \citep{Mateo:1998p773}. However, it is also clear that satellite galaxy formation continues after reionisation; three galaxies that end up as satellites of the Milky Way at $z=0$ are formed between $z=8$ and $z=6.7$.

\begin{figure}
\centerline{\includegraphics[width=1.0\hsize]{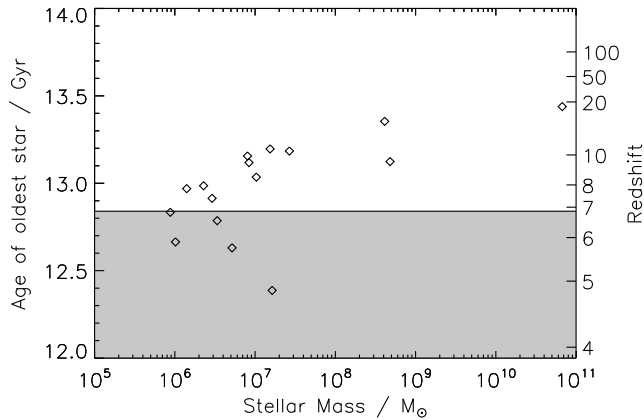}}
  \caption{Time of first star formation for satellite galaxies that survive to $z=0$ (including the MW itself) in the Reference run against their stellar mass at $z=0$. The region shaded in grey represents the lookback times which have not been simulated in the high resolution Cooling and Feedback runs. This illustrates that satellite galaxy formation is incomplete until $z=4.8$ (i.e. after a lookback time of 12.4 Gyr). Note that this plot only shows galaxies in the Reference run.}
\label{starages_masses}
\end{figure}

In fact, we find that the Milky Way satellite halo that began forming stars latest in the Reference run does so at $z=4.8$, or a lookback time of 12.4 Gyr, after reionisation is complete. This is illustrated on Fig.~\ref{starages_masses}, where we plot the stellar mass of each Milky Way satellite galaxy in the Reference run at $z=0$ against the age of their oldest star particle. As we note in section \ref{methods_halomaker}, we are unable to identify galaxies with a stellar mass below 3.5$\times10^5$\Msolar in the Reference run. Hence we cannot discount the possibility that the Cooling and Feedback runs might form more galaxies with lower masses that survive as Milky Way satellite galaxies. All we can conclude is that every galaxy above this mass threshold that survives as a Milky Way satellite in the Cooling and Feedback runs (through the twinning procedure) also survives as a satellite in the Reference run. However, it is possible that the trend of lower mass satellite galaxies forming at lower 
redshifts continues in the Cooling and Feedback runs, where stellar masses is better resolved.

\begin{figure*}
\centerline{\includegraphics[width=1.0\hsize]{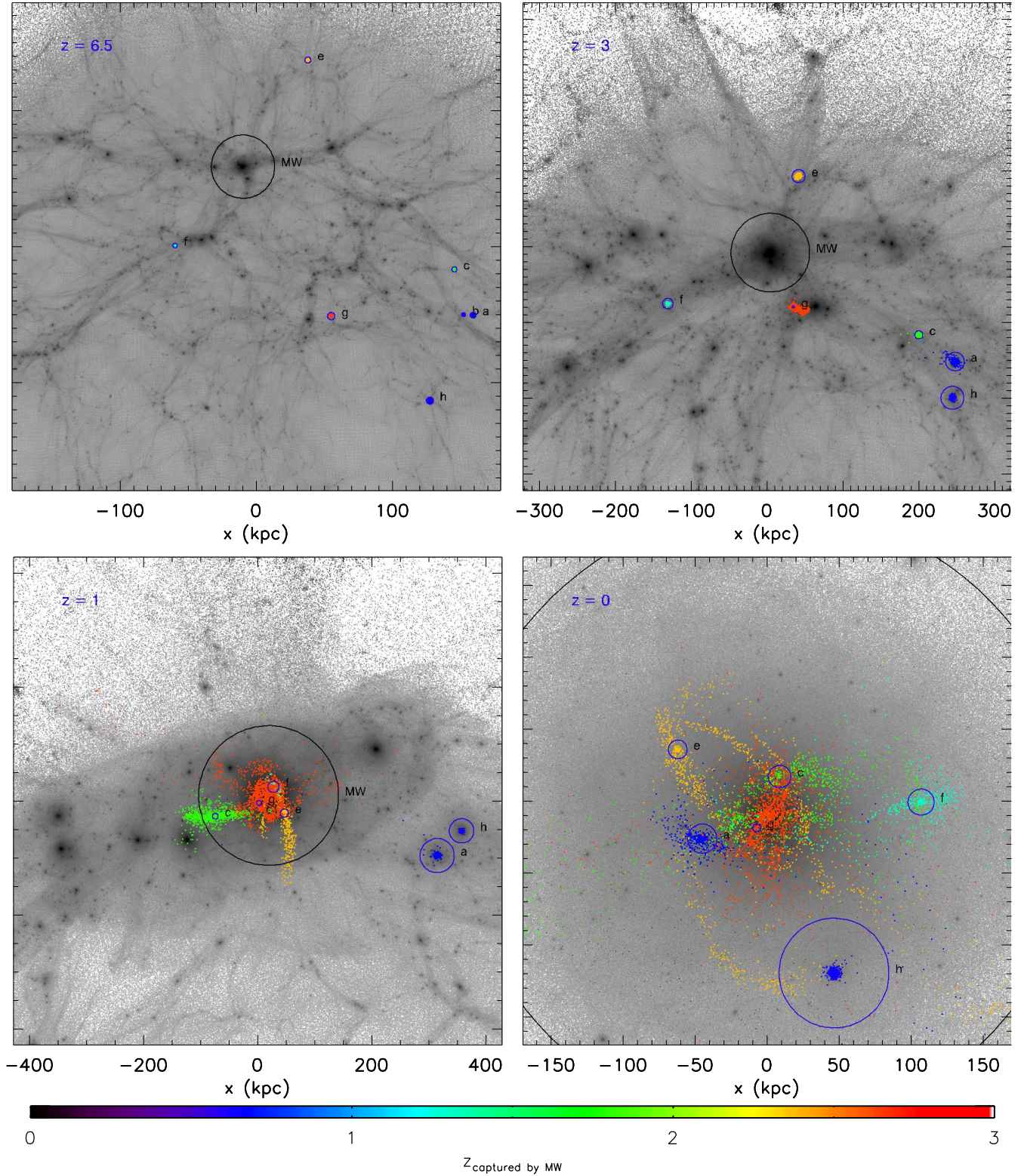}}
\caption{Tracking the location of dark matter particles in MW satellite progenitor halos from $z=6.7$ to $z=0$ in the Reference run. Each image is a projection of a cubic volume of length shown on the x-axis in physical kpc. Blue circles represent the virial radius of each halo tracked; the MW halo is shown as a black circle. The image at $z=0$ lies largely inside the Milky Way virial radius. The colour of the particles in a halo represents the redshift at which the halo is captured and becomes a subhalo of the MW (see the colour bar).  Halos captured before $z=1$ exhibit significantly more stripping at $z=0$ than halos captured after $z=1$. Halo `g' (in orange) is captured and partially stripped by another halo which, in turn, is captured and completely disrupted by the Milky Way between $z\leq3$ and $z=1$, while halo `g' itself survives as a luminous satellite galaxy of the MW at $z=0$.}
\label{dm_z9_to_z0}
\end{figure*}

Key to the survival process of satellite galaxies is the mass stripping they undergo as a function of time. We visualise this in Fig \ref{dm_z9_to_z0} where we identify which halos containing stars at $z=6.7$ survive to become Milky Way satellites at $z=0$ and follow their dark matter particles through cosmic time. We locate these particles in outputs of the simulation at $z=3, 1$ and $0$, and overplot them on top of their respective underlying density fields, colour coding them according to the redshift at which the halo they belong to is captured by that of the Milky Way. We find that halos captured before $z=1$ experience significant disruption despite the core of the halo surviving (halos `c', `e', 'f' and `g' in Fig.~\ref{dm_z9_to_z0}). (By contrast, we find that the only stars formed by $z=6.7$ that are stripped from their galaxies are from satellites captured before $z=3$). The surviving satellite that is captured just after $z=3$ (halo `g') arrives in the Milky Way halo as a subhalo of another halo, 
which is subsequently disrupted by the Milky Way. Hence this halo has already experienced some stripping by $z=3$, as shown in Fig. ~\ref{dm_z9_to_z0}.

\begin{figure*}
\centerline{\includegraphics[width=1.0\hsize]{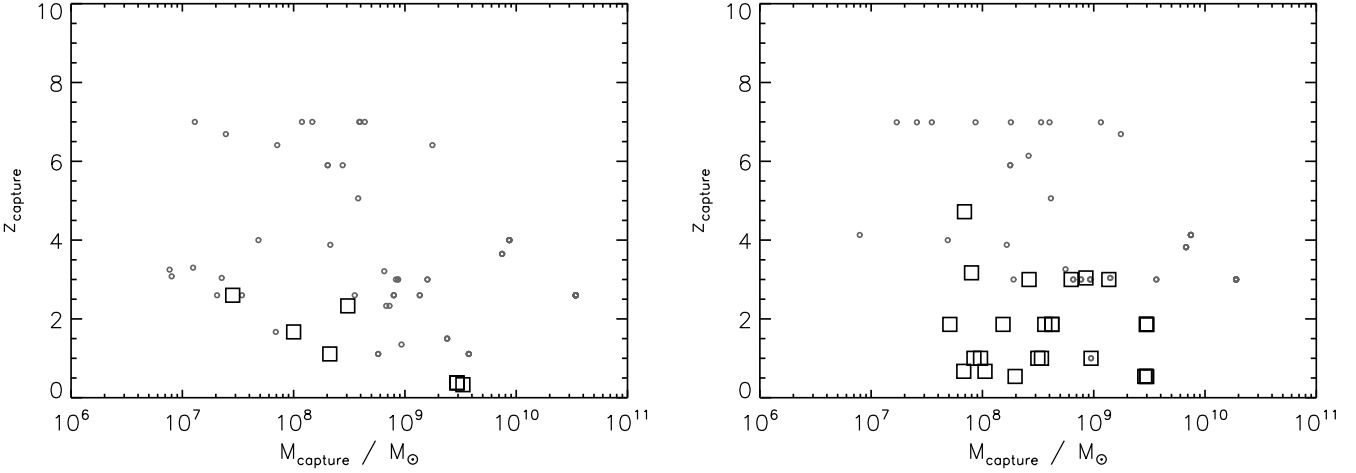}}
 \caption{Survival of halos as MW satellites at $z=0$. Both panels show the redshift at which a halo becomes a MW sub-halo, $z_{\rm capture}$, against the sub-halo mass when this capture happens, $M_{\rm capture}$; on the left is the Reference run, and on the right is the Adiabatic run. Black squares survive as MW satellite halos at $z=0$. Note that they can be completely disrupted as long as they become part of another MW satellite halo by $z=0$ in the disruption process. Grey circles are completely disrupted and simply become part of the diffuse MW halo by $z=0$. Halos captured by the MW host before $z=3$ do not survive to $z=0$, with one exception in the Adiabatic run. Less than half of the low-mass halos (M$_{\rm{vir}} \leq 10 ^9$ M$_\odot$) captured after $z=3$ are able to survive to $z=0$ in the Reference run, whereas the vast majority of them survive 
in the Adiabatic run.}
\label{survivalproperties}
\end{figure*}

In Fig.~\ref{survivalproperties}, we investigate satellite survival to $z=0$ in more detail. We compare the redshift at which a halo is captured by the Milky Way halo against its mass at capture in two simulations:
the Reference run and the matching Adiabatic run. The only difference between these two runs is that an extra right hand side 'sink' term is included in the energy equation of the gas in the Reference run to model
losses due to radiative cooling, as well as star formation (see section \ref{methods_numsim}). We find that out of all the halos which survive to $z=0$ in these two runs (6 in the Reference run, 20 in the Adiabatic run), only one is captured at $z > 3$ in the Adiabatic run and none in the 
Reference run. Moreover, higher mass halos (M$_{\rm{vir}} > 10^9 \sim$M$_\odot$) only survive if they are captured later so that the highest mass satellites at $z=0$ are systematically the ones that are captured last. It thus appears that the inclusion of radiative cooling in simulations of Milky Way-like galaxies has a dramatic impact on the survival of the satellite galaxies. We discuss the reasons for this discrepancy in section \ref{satstoday_physics}.

their pure dark matter counterparts. This is explained by two effects: (i) it is more difficult to strip mass from satellites as their central density increases and (ii) the cuspiness (and central density) of the host halo is increased. It is interesting to note that effect (i) could in principle lead to the opposite effect, i.e. an increase in the lifetime of the satellites as it makes them more concentrated and thus more resistant to tidal disruption, but we find that reduction in dynamical friction time scales due to mass increase dominates.  Of course these conclusions could be altered if a substantial mass of baryons was ejected out of the satellites, to the point where the trend that we measure could even be reversed.  \cite{Pontzen2012} argue that such a reversal is plausible, though it would require a significantly more efficient feedback mechanism than the one we observe in section \ref{feedback_sne}.

\begin{table*}
\begin{tabular}{l l l l l l l}
  \textbf{High z} & \textbf{Simulation} & \textbf{Total Number} & \textbf{Total Number} & \textbf{Satellite Mergers} & \textbf{Merged with MW}  \\
& & \textbf{high z (twinned)} & \textbf{z = 0} & \textbf{Destroyed (high z $\rightarrow$ z = 0)} & \textbf{(high z $\to$ z=0)}\\
 \hline
\multirow{2}{*}{11} & Cooling & 66 (36) & 1 & 0 (0 $\rightarrow$ 0) & 65\\
& Feedback & 65 (35) & 1 & 0 (0  $\rightarrow$ 0) & 64 \\
& Reference & 36 (36) & 0 & 0 (0  $\rightarrow$ 0) & 36 \\
  \hline
 \multirow{2}{*}{9} & Cooling & 91 (61) & 2 & 0 (0  $\rightarrow$ 0) & 89\\
& Feedback & 89 (60) & 2 & 0 (0  $\rightarrow$ 0) & 87\\
& Reference & 63 (63) & 1 & 0 (0  $\rightarrow$ 0) & 62 \\
  \hline
 \multirow{2}{*}{8.5} & Cooling & 89 (71) & 3 & 0 (0  $\rightarrow$ 0) & 86\\
& Feedback & 89 (71) & 3 & 0 (0 $\rightarrow$ 0) & 86\\
& Reference & 77 (77) & 2 & 0 (0  $\rightarrow$ 0) & 75\\
  \hline
 \multirow{2}{*}{8} & Cooling & 82 (68) & 3 & 1 (2  $\rightarrow$ 1) & 78\\
& Feedback & 90 (73) & 3 & 1 (2  $\rightarrow$ 1) & 86\\
& Reference & 85 (85) & 4 & 0 (0  $\rightarrow$ 0) & 81 \\
  \hline
 \multirow{2}{*}{6.7} & Cooling & 85 (78) & 6 & 1 (2  $\rightarrow$ 1) & 78\\
& Feedback & 85 (79) & 6 & 1 (2  $\rightarrow$ 1) & 78\\
& Reference & 107 (107) & 7 & 1 (2  $\rightarrow$ 1) & 99 \\
\end{tabular}
 \caption{Table of the fate of galaxies formed between $z=11$ and $z=6.7$. The 6 columns are, from left to right: (1) the redshift at which the stellar population is sampled in the Reference, Cooling and Feedback runs (`high z'); (2) the simulation name; (3) the number of galaxies which become satellites of the MW between the sampled redshift and $z=0$ not including the main MW progenitor halo (in brackets, the number of those galaxies whose twin halos in the Reference run also contain at least a galaxy); (4) the number of galaxies surviving as MW satellites at $z=0$; (5) the satellite progenitors destroyed by mergers with other satellite progenitors - the figures in brackets show the number of objects taking part in mergers at high z, followed by the resulting number of objects after the satellite progenitor-satellite progenitor mergers at $z=0$; (6) the number of galaxies merged with the MW and destroyed between high z and $z=0$. See section \ref{methods_tracking} for a description of how these numbers are 
calculated. We find that the large majority of the halos containing galaxies captured by the MW by $z=0$ merge with it and are destroyed, with two galaxies merging with each other before being captured by the MW and becoming a satellite galaxy. More mergers of MW progenitor galaxies are found, but these are all completely disrupted and destroyed after capture by the MW. We find that satellite galaxy formation in the Reference run is not complete by the lowest redshift reached by the high resolution runs (z=$6.7$). By looking at the ages of the star particles in satellite galaxies at $z=0$ in the Reference run, we find that satellite galaxy formation continues until at least $z=4.8$ (see Fig.~\ref{starages_masses}).}
\label{star_fate}
\end{table*}

\subsection{Dark Matter Satellite Halos}
\label{satstoday_dm}

\begin{figure}
\centerline{\includegraphics[width=1.0\hsize]{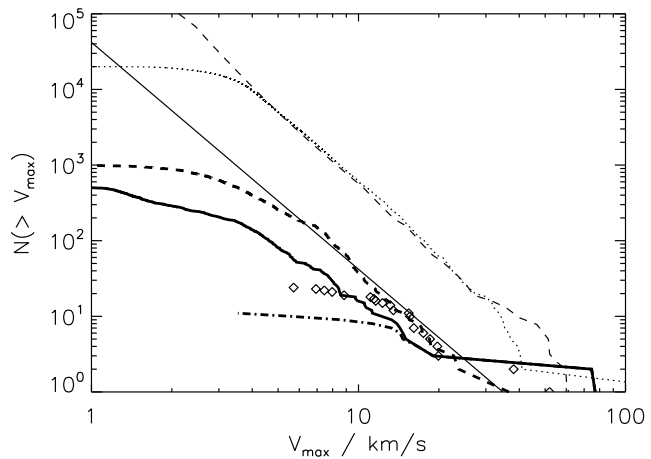}}
 \caption{Cumulative maximum circular velocity (V$_{max}$) functions comparing our results to high resolution dark matter only simulations of MW-like objects at $z=0$ published in the literature. The Reference run data is shown as a thick solid line, and the Dark Matter run as a thick dashed curve. Via Lactea II (VLII) data \protect\citep{Diemand:2008p986} is overplotted as a thin dotted curve. A fit to the Aquarius data \protect\citep{Springel:2008p979} is plotted as a thin dashed line. The empirical formula for N($<$V$_{max}$) given by \protect\cite{Reed:2005p2075} is plotted as a thin solid line. Observational data from \protect\cite{Mateo:1998p773,Bekki:2005p1869,Bekki2009,Wolf:2010p1868} is shown as diamonds. In each case, V$_{max}$ is given by max($\sqrt{GM(<r)/r}$), where M is the mass inside r$_{50}$, the radius at which the density exceeds 50$\rho_{crit}$.}
 \label{vmaxfn}
\end{figure}

In this section, we consider the population of dark matter subhalos of the Milky Way in our runs at $z=0$, comparing and contrasting their properties 
with similar dark matter simulations which exist in the literature.

For this purpose, we run a fourth simulation, the Dark Matter run, which is a pure dark matter version of the Reference run (see section \ref{methods_numsim}). The results of this simulation are described in more detail in section \ref{satstoday_physics}. We use it to compare our results directly with the Aquarius \citep{Springel:2008p979} and Via Lactea II \citep{Diemand:2008p986} simulations, which are the most resolved dark matter only simulations of MW-like 
objects available to date. The most basic comparison, a cumulative maximum circular velocity function, is presented in Fig. \ref{vmaxfn}. In this figure, we also overplot the empirical prescription proposed by 
\cite{Reed:2005p2075} for the Milky Way halo in our Dark Matter run at z=0. We find that our Dark Matter run satellite halo data is well represented by this prescription, but that the Reference run predicts significantly fewer  
satellites at the low velocity end (between a factor 2 and 3 for $V_{max} < 20 $km/s), and more massive satellites ($V_{max} > 30 $km/s). We discuss the impact of simulation physics on the maximum 
circular velocity function in the next section (\ref{satstoday_physics}).

From figure \ref{vmaxfn}, it is apparent that, whilst the cumulative maximum circular velocity function of our Dark Matter run has the same shape ($N(> V_{max}) \propto V_{max}^{-3}$) as that measured in both the Aquarius and Via Lactea II simulations, its
normalisation is more than an order of magnitude lower. This large discrepancy can almost entirely be attributed to our choice for the mass of the MW host halo since our agreement with the \cite{Reed:2005p2075} 
prescription is quite reasonable (better than 20 \%).  Our host halo has a V$_{max}$ of 126km/s in the Dark Matter run as opposed to $\sim$200km/s in the Aquarius or VLII simulations, i.e. 40\% less than either of these simulations. Note that in figure \ref{vmaxfn}, we plot all subhalos inside r$_{50}$ (denoting the radius at which the density is above 50$\rho_{crit}$) as opposed to r$_{200}$, which we use in the rest of this paper. This is to allow comparison with \cite{Reed:2005p2075,Diemand:2008p986,Springel:2008p979}. By contrast, the Reference run lies below the empirical formula, which we comment on in section \ref{satstoday_physics}. \cite{Springel:2008p979} report that their simulations overshoot the fitting formula of \cite{Reed:2005p2075} by a factor $\sim 3$ which they argue most likely 
arises from a systematic effect in the numerical technique used to perform the runs.  On the other hand, \cite{Madau:2008p2077} suggest it is due to the different (WMAP 1 instead of WMAP 3/5/7, i.e. different 
normalisation, $\sigma_8$, and/or tilt, $n_s$, of the power spectrum) cosmology the Aquarius simulations employ.  In any case, the remarkable conclusion that we 
draw from this comparison exercise, is that for a MW host halo with  V$_{max}$=126 km/s in the Dark Matter run as opposed to $\approx$ 210 km/s in the Aquarius or VLII simulations, i.e. a difference of 
about 65 \%, one gets a suppression in the number of satellites by about a factor of 10, which is enough to match the observed abundance of MW satellites with V$_{max}$ between 10 and 30 km/s. 
Now, there is still an ongoing debate as to what the exact mass of the Milky Way halo is (e.g. \cite{Battaglia:2005p1492,Karachentsev2005a,Watkins:2010p1876}). 
Our simulated halo admittedly lies at the very low end of the estimated range of values ($4.32 \times 10^{11} \rm{M}_\odot$ within 195 kpc), and Aquarius and Via Lactea II somewhat on the high side (1.85 $\times 10^{12}$ M$_\odot$ within 245 kpc). It is not the purpose of the present paper to constrain the Milky Way mass, but only to illustrate how the uncertainty in the mass of the Milky Way halo and the inclusion of baryonic physics translate into an uncertainty in the number and properties of MW satellites one predicts. Thus we simply remark that the observed abundance of these satellites, taken at face value, 
seems to favour a less massive MW halo.

\subsection{The Effect of Baryonic Physics on Satellite Galaxy Survival}
\label{satstoday_physics}

\begin{table*}
\begin{tabular}{l l l l l l l}
   \textbf{Redshift} & \textbf{Halo Type} & \textbf{Dark Matter} & \textbf{Adiabatic} & \textbf{Reference} & \textbf{Cooling} & \textbf{Feedback} \\
  \hline
 \multirow{2}{*}{0} & Independent Halos & 2084 & 1706 & 1731 & - & -\\
                                   & Subhalos & 2730 & 2418 & 2209 & - & -\\
   \hline
  \multirow{2}{*}{7}  & Independent Halos & 2405 & 1997 & 1997 & 2142 & 2104\\
                                   & Subhalos & 598 & 449 & 343 & 445 & 390\\
\end{tabular}
  \caption{Total number of independent halos and subhalos (dark and luminous) in the Dark Matter, Adiabatic and Reference runs at two sampled redshifts. Runs containing baryons compare favourably, whereas the Dark Matter run contains significantly more halos for a given redshift. There are also fewer subhalos in the Reference run than either of the other two runs, as explained in section \ref{satstoday_physics}.}
\label{physics_halonumber} 
\end{table*}

\begin{figure}
\centerline{\includegraphics[width=1.0\hsize]{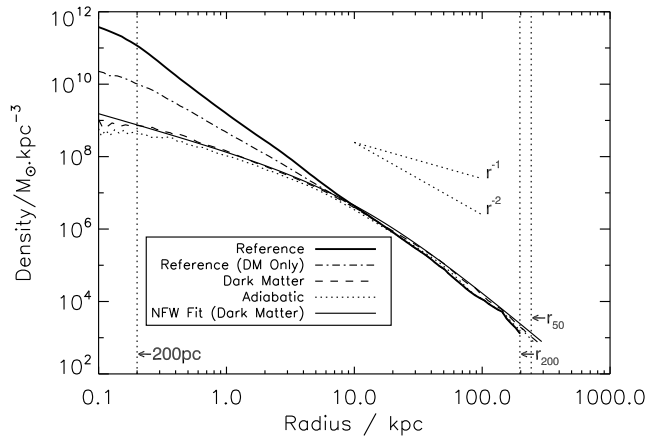}}
 \caption{Density profiles of the main Milky Way-like halo up to r\Subvir at $z=0$ for the Reference (thick solid line), Adiabatic (dotted line) and Dark Matter (dashed line) runs. We plot total density, i.e. dark matter, gas and stars in the Reference run and dark matter and gas in the Adiabatic run; the dot-dashed line represents the DM density in the Reference run. We overplot a vertical dotted line at 200pc, equivalent to AMR level 16 at $z=0$, i.e. 2 levels below the highest resolution reached by the runs to indicate the scale below which the gravitational force is underestimated, and lines at r$_{200}$, used in the bulk of the text and r$_{50}$, used for comparison purposes with figure \ref{vmaxfn}. Power law profiles scaling like r$^{-1}$ and r$^{-2}$ and a NFW profile fit to the DM run (thin solid line) are also overplotted.}
\label{densityprofile}
\end{figure}

\begin{figure}
\centerline{\includegraphics[width=1.0\hsize]{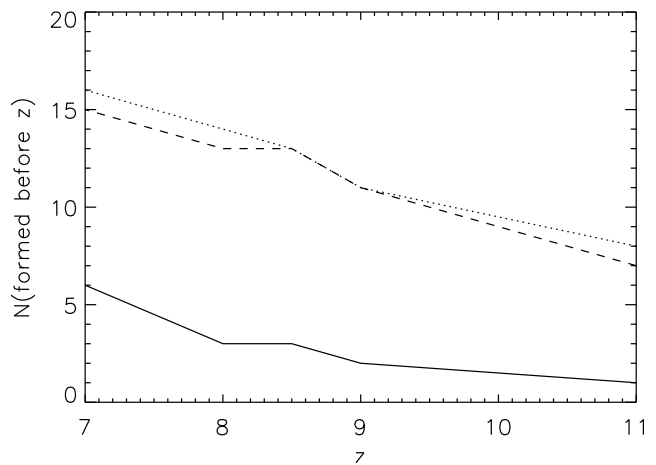}}
 \caption{Number of satellite galaxies at $z=0$ formed at or before a given redshift. Satellite galaxies are defined as sub-halos whose progenitors at high redshift have a twin in the Feedback run that contains one or more galaxies (see section \ref{satstoday_tracking}). Each line represents a different run used to track halos down to $z=0$. The solid line shows the Reference run, the dashed line shows the Dark Matter run and the dotted line shows the Adiabatic Run. The inclusion of gas cooling reduces the number of galaxies at high redshift that survive down to $z=0$ by a factor $\sim$ 3 to 10 depending on redshift. As discussed in the text, the central density profile of the Milky Way at $z=0$ plays a key role in governing the survival of satellite galaxies.}
\label{fategraph}
\end{figure}

As described in section \ref{methods_numsim}, we run three simulations to $z=0$, the Dark Matter Run, the Adiabatic Run and the Reference Run. We also follow galaxies formed by $z=6.7$ in the Cooling and Reference run to $z=0$ by using the Reference run's merger tree. Only the Reference Run includes star formation up to z=0. The first thing to note is that the Dark Matter run produces more (sub-)halos ($\approx 20 \%$ more) than any of the runs containing baryons; we present exact numbers in table \ref{physics_halonumber}. These results are found both at z=7 and z=0. It thus appears that pure dark matter simulations, which do not include baryonic pressure forces, allow for the more efficient collapse of halos than simulations that contain baryons.

Secondly, as described earlier in section \ref{satstoday_tracking}, we find fewer subhalos in the Reference run than in the two other runs. In figure \ref{fategraph}, we quantify this effect by plotting the number of high redshift galaxies surviving to become satellite galaxies of the Milky Way at $z=0$ (see sections \ref{methods_tracking} and \ref{satstoday_tracking} for details on the tracking of satellites from high to low redshift and table \ref{star_fate} for numbers specific to the Reference run). We find that the Reference run contains far fewer (factor 3 to 7 depending on redshift) surviving satellite galaxy sub-halo hosts than the Dark Matter or Adiabatic runs. Similarly, figure \ref{survivalproperties} shows that while in the Adiabatic run, all halos captured by the Milky Way after $z=3$ survive to $z=0$, in the Reference run only the less massive halos survive to $z=0$. Finally, we note that in figure \ref{vmaxfn}, the Reference run subhalo cumulative maximum circular velocity function 
at $z=0$ lies below the Dark Matter run's. There is hence a body of evidence to suggest that including gas cooling in our simulations of a Milky Way-like halo has significantly altered the population of halos, both dark and luminous, surviving to $z=0$.

This discrepancy can be explained as the result of increased dynamical friction on the more massive infalling satellites, which is exacerbated by differences between the simulations in the density profile of the Milky Way. Indeed, the Chandrasekhar formula, which correctly encapsulates the basic physics of dynamical friction, states that the friction force is proportional to the mass of that object and on the density of the surrounding medium \citep{Binney:2008p1799}. Figure \ref{densityprofile} shows that the total (gas \& DM) density profiles of the main MW halo at $z=0$ are very similar between the DM and Adiabatic runs, with the Adiabatic run having a slightly lower density overall. However, the profiles diverge at around 10kpc; at a radius of 2kpc, the density in the Reference run is several times higher than in the other runs. This is due to the fact that gas in the Reference run is able to cool, and hence the density of the halo within a radius of 10kpc is significantly higher in this run, 
since cooled baryons are able to condense at the centre, pulling DM along with them \cite{Flores1994}.

We analyse the halos containing satellite galaxies that survive to $z=0$ in the Adiabatic run but not the Reference run and find the following. All of these halos pass close to the Milky Way's centre in both the Reference and Adiabatic runs. In addition, when close to the centre, the halos' velocities are highly radial, suggesting that these subhalos have experienced sufficient dynamical friction and lose most of their angular momentum. Depending on their mass and density, the satellites survive 1 to 7 passes before their orbit decays to the point where they merge completely with the Milky Way halo in the Reference run. In the Adiabatic run, where the central density of the Milky Way is much lower, the orbits decay much more slowly, and as a result these objects are not destroyed by $z=0$. By comparison, the surviving satellites in the Reference run do not pass close to the centre of the Milky Way, and thus survive to $z=0$.

To conclude, even if our sub-halos had exactly the same mass in the DM and Reference runs, a higher central density of the host halo   
would cause their orbits to decay faster, thus decreasing their survival time, as found by \cite{RomanoDiaz:2010p2098} and \cite{Schewtschenko:2011p2097}. This effect is exacerbated as the satellites in the Reference run are also more concentrated, rendering them less prone to tidal disruption at large radii, retaining more mass on their way to the centre of the host. This higher mass makes them more susceptible to dynamical friction. It should be noted that we use an AMR code, as opposed to SPH as previous authors do. One consequence of this is that gravitational force resolution is dependent on the resolution of the grid structure. However, due to our relatively high resolution (50pc in the Reference run), our force resolution in the centres of the MW halo and its satellites is competitive with, if not better than, that of previous authors.
We thus believe that the main caveat of our Reference run is that it does not include a feedback model, although we point out that standard supernovae feedback, as  modelled in our high-resolution Feedback run, is extremely unlikely to reverse the situation significantly. However,   
 \cite{Pontzen2012} have recently argued that the injection of energy from supernovae in the centre of dwarf galaxies 
 at $4>z>2$ can dramatically alter their halo density profiles. Whilst we are not able to confirm this effect with our own set of simulations, should 
it prove to be able to lower the central density of the Milky Way halo as well, we predict that more satellites will survive to $z=0$. However, whether this number 
will match that measured in pure DM simulations or will still reflect a significant suppression of satellites is likely to depend on the details of the numerical implementation of the feedback processes.

\subsection{Properties of Satellite Galaxies at $z=0$}
\label{satstoday_z0}

\begin{figure*}
\centerline{\includegraphics[width=1.0\hsize]{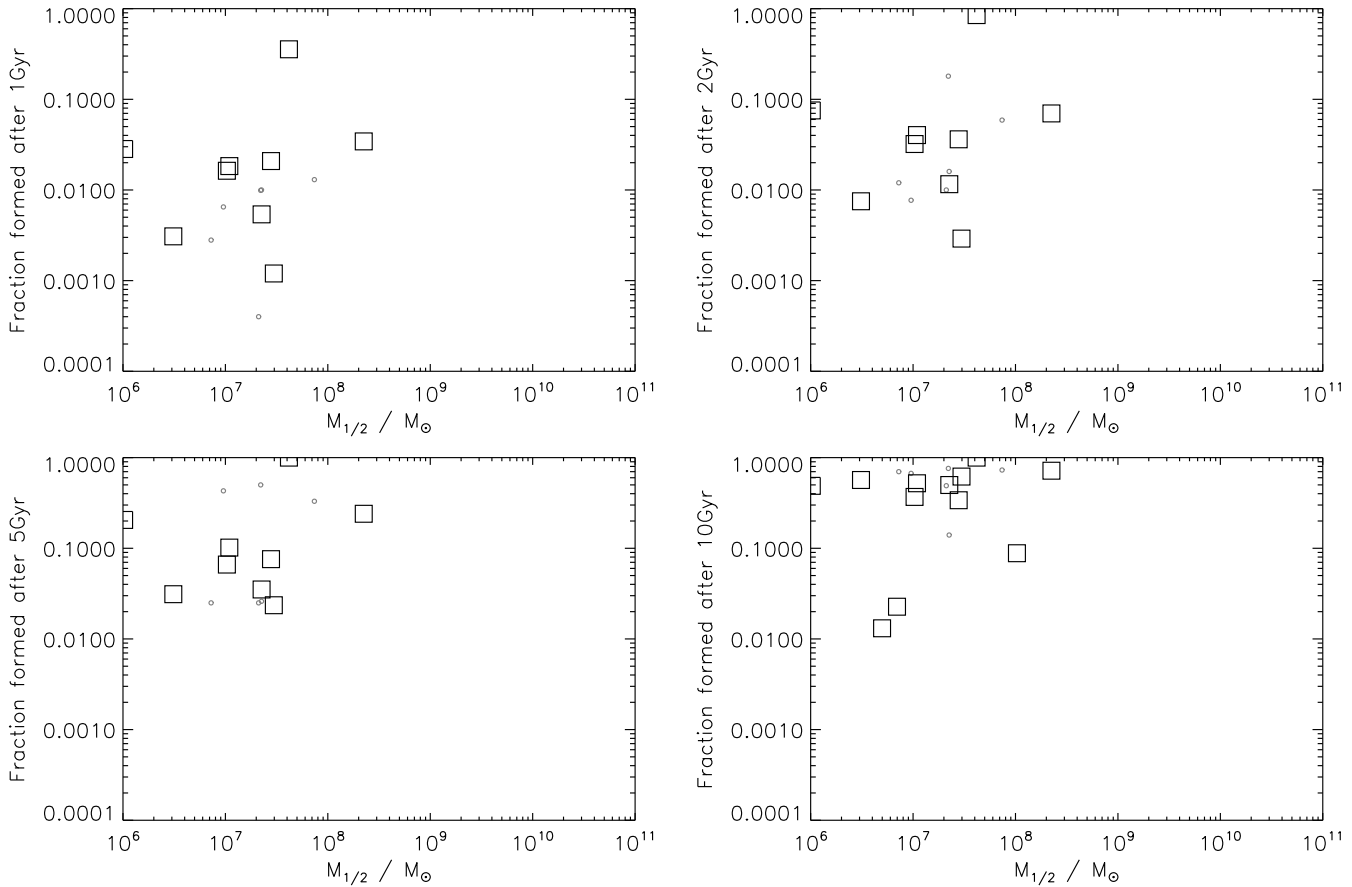}}
\caption{Star formation histories of satellite galaxies in the Reference run (black squares) compared to values deduced from observations of Milky Way satellites (\protect\cite{Orban:2008p1763} and \protect\cite{Wolf:2010p1868}: grey circles). Each panel shows the fraction of stellar mass formed between $z=0$ and the lookback time indicated on 
ordinate axis against the half stellar mass of the galaxy at $z=0$. Stellar fractions and masses in each plot match well with the values estimated from observations, suggesting that the star formation histories of satellite galaxies in the Reference run do not wildly differ from those of observed Milky Way satellites.}
\label{stellarmasscomparison}
\end{figure*}

\begin{figure}
\centerline{\includegraphics[width=1.0\hsize]{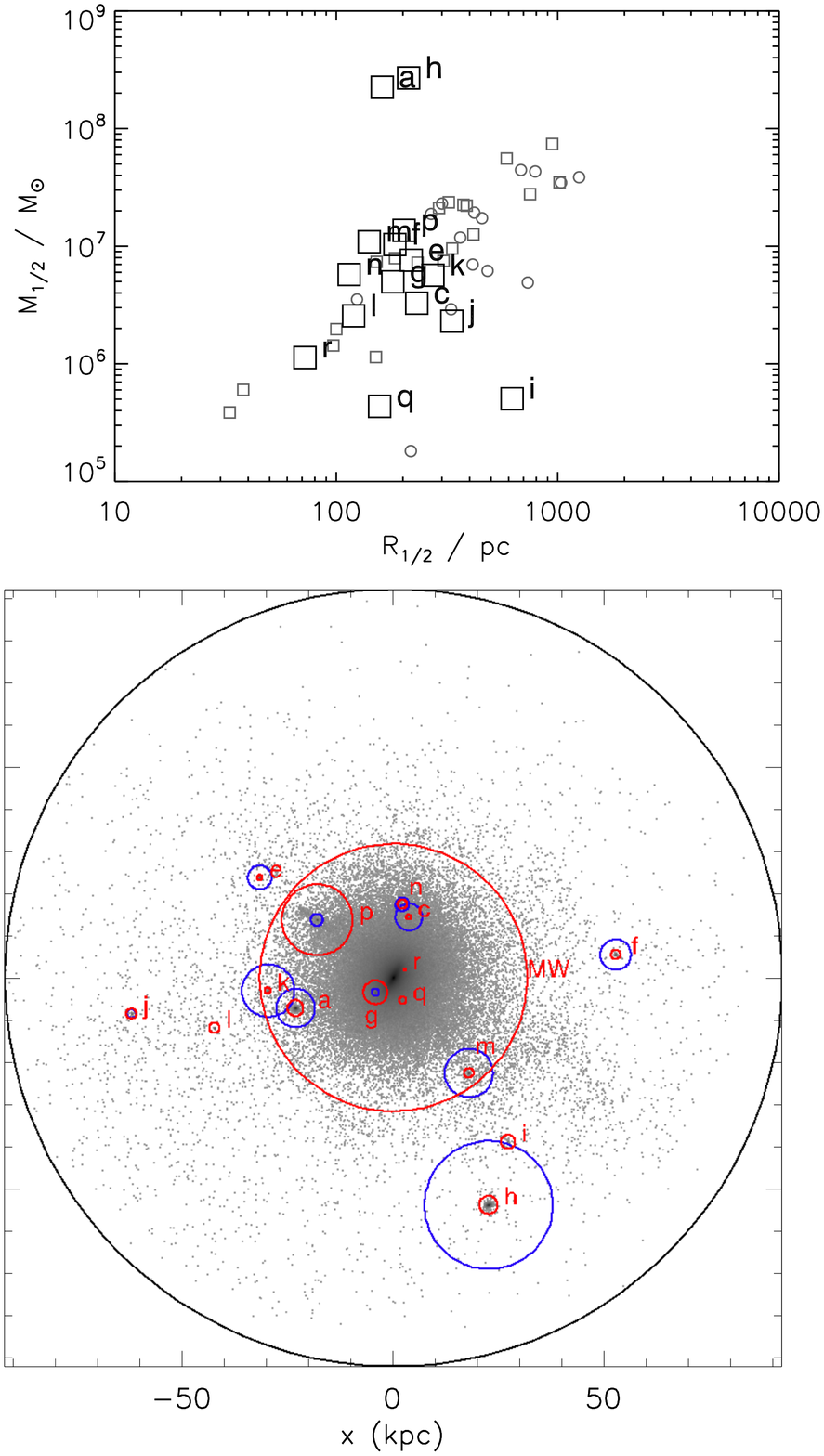}}
 \caption{Properties of satellite galaxies surviving down to $z=0$. The top panel shows half stellar mass against half-stellar-mass radius for satellite galaxies in the Reference run at $z=0$
 (large black squares), compared to half-light radii calculated for satellites of the Milky Way and M31 (small grey squares and small grey circles respectively). Data for the satellites of the Milky Way are taken from \protect\cite{Wolf:2010p1868}. Data for the satellites of M31 are derived from \AndromedaReferences. The bottom panel shows the spatial location of the satellite galaxies and their host DM sub-halos inside the Milky Way virial radius (red circles indicate galaxy radii, while blue circles represent sub-halo radii). The black circle shows the Milky Way virial radius. The background image is a cubic projection of the stellar density field. The two satellite galaxies (labelled $a$ and $h$) which lie above the observed data in the top panel are galaxies that have been little affected by stripping due to their late capture by the Milky Way host. Note that the Milky Way data does not include the Magellanic clouds or ultrafaint satellites.}
\label{massfn_stars}
\end{figure}

In this section we compare the properties of galaxies found at $z=0$ in our simulations to observations of known satellite galaxies of the Milky Way and M31. To do this, we make use of the Reference run results measured at $z=0$ directly rather than relying on the satellite tracking algorithm described in section \ref{satstoday_tracking}. In this run, we find fifteen satellite galaxies using a direct galaxy identification method described in section \ref{methods_halomaker}. Four of these (labelled $i$, $l$, $r$ and $q$ on Fig.~\ref{massfn_stars}) do not have associated dark matter host halos identified by our halo detection algorithm. Three of these, $i$, $l$ and $q$, once had hosts that fell below our halo mass resolution threshold of $2.2 \times 10^6$ M$_\odot$ by z=0. The host of the fourth, $r$, merged with another halo at z=5.3, which then merged with the Milky Way at z=2.3. We would thus require a higher dark matter resolution to comment on whether satellite galaxies are capable of losing their 
dark matter host halos without themselves being destroyed by stripping. We further note that including a DM sub-halo host with a mass just below our mass resolution threshold ($2.2 \times 10^6$ M$_\odot$) would not affect the results we present here.

Figure \ref{massfn_stars} shows a plot comparing the half stellar mass against half-light radius of our simulated satellite galaxies (grey squares), to the satellite galaxies of the Milky Way (from \cite{Wolf:2010p1868}) (white circles) and M31 (from \AndromedaReferences) (white squares).  The population of simulated galaxies lie in the same region as the observed satellites except for the two galaxies labelled $a$ and $h$ on the figure, which lie well above the observational data. These are the most massive galaxies in our sample, which have recently been captured by the Milky Way host halo and thus have not experienced significant stripping by $z=0$. We note that beyond their larger stellar half mass, which can be 
somewhat overlooked since the data from \cite{Wolf:2010p1868} do not contain the LMC or SMC, these simulated galaxies seem too compact when compared to the rather 
well defined observational relation linking satellites size and mass. While resolution undoubtedly plays an important role in getting accurate estimates of the sizes of simulated 
objects,  it is nevertheless striking that satellite galaxies which lie on top of the observed results all exhibit large tidal tails (figure~\ref{dm_z9_to_z0}) and thus evidence of tidal stripping, compared to the two most massive galaxies (`a' and `h' in figure~\ref{dm_z9_to_z0}), which enter the Milky Way halo much later, after z=1. This is relevant because the LMC and SMC are believed to have entered the Milky Way later than the other satellite galaxies \citep{Besla2010}, which can explain the morphological differences between them.

Moreover, when we compare star formation histories of satellite galaxies in the Reference run with those derived from the analysis of colour-magnitude diagrams (CMDs) of 
Milky Way satellite galaxies observed by \cite{Orban:2008p1763} we find reasonable agreement. This is demonstrated in Fig.~\ref{stellarmasscomparison}, where we plot the fraction of stars formed after lookback times of 1Gyr, 2Gyr, 5Gyr and 10Gyr both in simulated (grey squares) and observed (white circles) galaxies. It is somewhat reassuring that the star formation histories of our simulated satellite galaxies are broadly correct since we have
previously shown (Fig~\ref{starages_compare}) that they are fairly independent of resolution. Perhaps more importantly, this agreement also suggests 
that feedback, whatever its form and origin, cannot drastically alter the star formation histories of these galaxies: models where significant feedback at $z>1$ completely quenches 
late star formation are clearly ruled out by the observational data. This makes it all the more challenging for stellar feedback to soften 
cusps of dark matter sub-halos and certainly favours a rapid, irreversible mechanism, very localised in time such as the one suggested by 
\cite{Pontzen2012}.

\section{Discussion and Conclusions}
\label{discussion}

The work discussed in this paper has made use of the \Simname suite of high (a few tens of  parsec) to ultra-high (sub-parsec) resolution cosmological hydrodynamic re-simulations to investigate the effect of baryonic physics on the evolution of a `Milky Way' and its satellite galaxies.  Whilst various other authors have simulated satellites of Milky Way-like galaxies down to $z=0$, ours are the first to reach sub-parsec resolution down to the end of the epoch of reionisation. The motivation for this was to analyse in detail the evolution of satellite galaxies around the epoch of reionisation, which has been posited as a mechanism for suppressing star and galaxy formation in dwarf halos and hence for shaping the population of satellite galaxies that we observe around the Milky Way and M31. To the best of our knowledge, we are also the first authors to use an adaptive mesh refinement (AMR)  technique to study the evolution of these satellite galaxies down to $z=0$ (other studies thus far either used smoothed 
particle hydrodynamics (SPH), or ended their simulations at higher redshifts).
 
In agreement with e.g. \cite{Guo2010a,Ricotti:2005p2099,Wadepuhl2010a}, we find that reionisation appears not to efficiently stop star, or even galaxy formation. Instead, we find that satellite galaxy formation continues down to at least  $z=4.8$ in our lowest ($\sim$50 pc) resolution simulation. The number of luminous satellite galaxies formed before reionisation ($z=8.5$ in our case) is found to be far lower than the number of observed Milky Way satellites. These results are consistent with e.g. \cite{Okamoto:2009p1754,Hoeft:2006p1811}: like these authors, we find that, down to the end of the reionisation era, there exists a threshold in v$_{max}$ of about 10 km/s below which halos remain dark, never forming stars. This threshold persists at later redshifts, i.e. well after reionisation has ended. This is probably due to the fact that efficient atomic gas cooling increases halo central densities and hence v$_{max}$, separating halos that can cool gas and form stars from those that cannot. There are, 
however, at least two major limitations in our work that prevent us from commenting further. First, we do not run a simulation with self-consistent star formation and ionisation, and hence we cannot quantify the precise effect of UV photoionisation on star formation in galaxies already forming stars. Secondly, our model of reionisation consists of a uniform background which neglects the effect of gas self-shielding from external photoionisation sources. Future studies which include ionising photon radiative transfer and hence self-consistent re-ionisation, should allow us to determine whether proximity effects and/or self-shielding significantly alter our conclusions, but this does not seem likely.

The effect of supernova feedback on the gas and stellar mass of high redshift (both pre- and post-reionisation) galaxies seems to be quite stochastic.
 Indeed, our results exhibit a large scatter, even though one might argue we detect a slight systematic trend of star formation being enhanced by feedback as 
galaxy mass increases. A key feature of our model of supernova feedback is that it consists of star particles (with a standard Salpeter IMF) injecting mass, energy, momentum 
and metals into the surrounding gas according to a Sedov blast wave solution deposited onto the grid 10Myr after they have formed. Together with our sub-parsec resolution, this means that we should be able to track the effect of supernovae 
explosions fairly realistically on scales typical of small molecular clouds. Outflows and heating from these supernovae reduce the amount of gas available for star formation, but blast wave compression and an excess of radiative cooling due to the injection of metals into the ISM can potentially increase it. Our results show that, on average, adding supernovae feedback does reduce the gas and stellar mass of galaxies in halos below $10^9$\Msolar (negative feedback)  but increases it for galaxies hosted by halos above $10^9$\Msolar, where the deeper potential and extra metal injection negates the impact of outflows (positive feedback). In any case, the stellar masses of individual galaxies are only changed by maximum 10-20 percent either way by supernova feedback, and one has to invoke a much larger energetic input  (for example a top-heavy IMF, or a large fraction of hypernovae)
 and/or one that is impervious to radiative losses (e.g. some kind of 'turbulent' energy) to overturn the situation in favour of negative feedback.

When we use lower resolution (50 pc) runs to track the descendants of galaxies in the high resolution (0.5 pc) Cooling and Feedback runs 
to $z=0$, we find that very few of the galaxies which are captured by the Milky Way progenitor halo survive to $z=0$. In fact, independent of the run used to perform the tracking (PureDM, Adiabatic or Reference), no satellite galaxy captured by the Milky Way progenitor before $z=3$ survives as Milky Way satellite at $z=0$. However, the main difference is that {\em all} galaxies (sub-halos) captured {\em after} $z=3$ in the Pure DM or Adiabatic runs survive, while only a small fraction 
of these does so in the Reference run. This is caused by the much higher central density of the Milky Way halo (and sub-halos) in the Reference run (the only one to host a MW galaxy), which significantly shortens the dynamical friction timescales for satellites to spiral to the centre of the halo and experience disruption. \cite{Libeskind:2010p1751} suggest that satellites in simulations including gas cooling (rather than pure dark matter) experience lower mass loss, although they are more radially concentrated. However, \cite{RomanoDiaz:2010p2098} find, as we do, that including baryon physics reduces the survival time of satellites. \cite{Schewtschenko:2011p2097} explains this discrepancy by noting that \cite{Libeskind:2010p1751} do not measure satellite survival but rather mass loss over time, whereas \cite{RomanoDiaz:2010p2098} and \cite{Schewtschenko:2011p2097} find that it is in the centre of the halo that the satellites experience the most mass loss, and hence the centre of the host is 
where the survival of satellites is determined.

We believe that the work presented here, which makes use of a completely different simulation technique and improves on the resolution of these previous studies, sheds a useful light on the issue. In particular, we emphasize that when we compare properties of the remaining satellite galaxies at $z=0$ in the Reference run to their observational equivalents for the Milky Way or M31, we find that the star formation histories, stellar masses and radii are in reasonable agreement. Only the more massive, recently captured satellites are found to be too compact when compared to Milky Way and M31 dwarf spheroidal satellites. This puts rather tight constraints on the feedback mechanisms (amount of energy, duration, timing) required to soften the cusps of dark matter halos as they cannot have a major impact on these properties.

Finally, beyond the importance of the role played by baryonic physics in determining the number of satellites of MW class halos, it is worth noting the extreme sensitivity of this number to the circular velocity of the host halo.  Whilst all simulated CDM halos in the literature match the shape of the $N_\mathrm{sat} \propto V_{max}^3$ relation from \cite{Reed:2005p2075}, not all of them agree on the constant of normalisation. Indeed, our Pure DM run agrees with the normalisation given by \cite{Reed:2005p2075} at the 10-20 \% level whereas the Aquarius \citep{Springel:2008p979} simulations overshoot it by a factor $\sim 3$, and by about 30 \% more than the Via Lactea II \citep{Diemand:2008p986} simulation.
Curiously enough, the cause of this significant discrepancy is still unresolved, even though the debate has received a lot of attention lately in papers such as \cite{Vera-Ciro2011,Boylan-Kolchin2012,WangJie2012}. Irrespective of this disagreement, our results, especially when baryonic physics is taken into account, argue in favour of a less massive (5 $\times 10^{11}$ M$_\odot  < $ M$_{vir} < 10^{12}$ M$_\odot$) MW halo. Note that if the likelihood for a satellite galaxy to survive down to $z=0$ is suppressed by even half as much as we find in this work when feedback is properly incorporated, models relying on N-body dark matter-only simulations, such as semi-analytic models (SAMs) should be revised accordingly.

\section{Acknowlegements}
\label{acknowledgements}
The authors would like to thank the anonymous referee for their detailed and helpful comments, which greatly helped to improve this work. We would also like to thank Romain Teyssier and Taysun Kimm for useful comments and discussions during the production of this paper. The simulations 
presented here have run partly on the JADE supercomputer at the {\em Centre Informatique National de l'Enseignement Sup\'erieure} (on resources allocated to project number GEN2191), and 
partly on the DiRAC facility jointly funded by STFC, the Large Facilities Capital
Fund of BIS and the University of Oxford.
We warmly thank Jonathan Patterson for his phenomenal help in using the latter. The authors would also like to thank Michelle Collins and Mike Irwin for allowing them to use their collated M31 satellite data, and Dylan Tweed for making his merger tree code available. We also acknowledge the Royal Astronomical Society for funding in support of this research. SG is funded by a STFC studentship and a Cosmocomp fellowship. JD and
AS's research is supported by Adrian Beecroft, the Oxford Martin
School and the STFC.

 \bibliographystyle{mn2e}
 \bibliography{satellitepaper}

\end{document}